\RequirePackage{fix-cm}
\documentclass{article}       

\usepackage{arxiv}
\usepackage{physics} 
\usepackage{siunitx} 
\usepackage{enumerate} 
\usepackage{subcaption}
\usepackage{multirow}
\usepackage{float,lscape}
\usepackage{pgfplots}
\usepackage{pgfplotstable}
\usepackage{tikz,pgfplots}
\usepackage{amsmath}  
\usepackage{amssymb}
\usepackage[misc,geometry]{ifsym} 
\usepackage{footmisc}
\usepackage{doi}

\newcolumntype{P}[1]{>{\centering\arraybackslash}p{#1}}
\makeatletter
\newcommand\footnoteref[1]{\protected@xdef\@thefnmark{\ref{#1}}\@footnotemark}
\makeatother

\begin{document}

\title{Using altmetrics for detecting impactful research in quasi-zero-day time-windows: the case of COVID-19} 

\author{Erik Boetto \\
DIBINEM, University of Bologna, Italy \\ 
E-mail: \texttt{erik.boetto@gmail.com}\\
ORCID: 0000-0002-2509-159X
\And 
Maria Pia Fantini \\
DIBINEM, University of Bologna, Italy \\
E-mail: \texttt{mariapia.fantini@unibo.it}\\
ORCID: 0000-0002-3257-6552
\And
Aldo Gangemi \\
STLab, ISTC-CNR, Rome, Italy \\
E-mail: \texttt{aldo.gangemi@cnr.it} \\
ORCiD: 0000-0001-5568-2684
\And 
Davide Golinelli \\
DIBINEM, University of Bologna, Italy \\
E-mail: \texttt{davide.golinelli@unibo.it}\\
ORCID: 0000-0001-7331-9520
\And
Manfredi Greco \\ 
DIBINEM, University of Bologna, Italy \\
E-mail: \texttt{manfredi.greco@studio.unibo.it}\\
ORCID: 0000-0001-6886-0160
\And 
Andrea Giovanni Nuzzolese\textsuperscript{ \Letter } \thanks{A.G. Nuzzolese led the work presented in this paper. The authors are sorted alphabetically as they equally contributed to this work.} \\
STLab, ISTC-CNR, Rome, Italy \\
E-mail: \texttt{andrea.nuzzolese@cnr.it} \\
ORCiD: 0000-0003-2928-9496
\And 
Valentina Presutti \\
LILEC, University of Bologna, Italy \\
E-mail: \texttt{valentina.presutti@unibo.it} \\
ORCiD: 0000-0002-9380-5160
\And 
Flavia Rallo \\
DIBINEM, University of Bologna, Italy \\
E-mail: flavia.rallo@studio.unibo.it
}

\maketitle

\begin{abstract} ---
On December 31st 2019, the World Health Organization (WHO) China Country Office was informed of cases of pneumonia of unknown etiology detected in Wuhan City. The cause of the syndrome was a new type of coronavirus isolated on January 7th 2020 and named Severe Acute Respiratory Syndrome CoronaVirus 2 (SARS-CoV-2). SARS-CoV-2 is the cause of the coronavirus disease 2019 (COVID-19). Since January 2020 an ever increasing number of scientific works related to the new pathogen have appeared in literature. Identifying  relevant  research outcomes at very early stages is challenging. In this work we use COVID-19 as a use-case for investigating: (i) which tools and frameworks are mostly used for early scholarly communication; (ii) to what extent altmetrics can be used to identify potential impactful research in tight (i.e. {\em quasi-zero-day}) time-windows. A literature review with rigorous eligibility criteria is performed for gathering a sample composed of scientific papers about SARS-CoV-2/COVID-19 appeared in literature in the tight time-window ranging from January 15th 2020 to February 24th 2020. This sample is used for building a knowledge graph that represents the knowledge about papers and indicators formally. This knowledge graph feeds a data analysis process which is applied for experimenting with altmetrics as impact indicators. We find moderate correlation among traditional citation count, citations on social media, and mentions on news and blogs. Additionally, correlation coefficients are not inflated by indicators associated with zero values, which are quite common at very early stages after an article has been published. This suggests there is a common intended meaning of the citational acts associated with aforementioned indicators. Then, we define a method, i.e. the {\em Comprehensive Impact Score} (CIS), that harmonises different indicators for providing a multi-dimensional impact indicator. CIS shows promising results as a tool for selecting relevant papers even in a tight time-window. Our results foster the development of automated frameworks aimed at helping the scientific community in identifying relevant work even in case of limited literature and observation time.
\end{abstract}
\section{Introduction}
\label{sec:intro}
A {\em zero-day} attack is a cyber attack exploiting a vulnerability (i.e. zero-day vulnerability) of a computer-software that is either unknown or it has not been disclosed publicly~\cite{Bilge2012}. There is almost no defense against a zero-day attack. In fact, according to~\cite{Bilge2012}, while the vulnerability remains unknown, the software affected cannot be patched and anti-virus products cannot detect the attack through signature-based scanning. 

On December 31st 2019, the World Health Organization (WHO) China Country Office was informed of cases of pneumonia of unknown etiology detected in Wuhan City (Hubei Province, China), possibly associated with exposures in a seafood wholesale market in the same city\footnote{WHO - World Health Organization. Situation report - 1, Novel Coronavirus (2019-nCoV), January 21st 2020 \url{https://www.who.int/docs/default-source/coronaviruse/situation-reports/20200121-sitrep-1-2019-ncov.pdf?sfvrsn=20a99c10_4}.}. The cause of the syndrome was a new type of coronavirus isolated on January 7th 2020 and named Severe Acute Respiratory Syndrome CoronaVirus 2 (SARS-CoV-2). Formerly known as the 2019 novel coronavirus (2019-nCoV), SARS-CoV-2 is a positive-sense single-stranded RNA virus that is contagious among humans and is the cause of the coronavirus disease 2019, hereinafter referred to as COVID-19~\cite{Gorbalenya2020}. Borrowing cyber security terminology, COVID-19 is a zero-day attack where the target system is the human immune system and the attacker is SARS-CoV-2. The human immune system has no specific defense against SARS-CoV-2. Being SARS-CoV-2 a new type of virus, there is no immunity provided by either natural or artificial immunity (i.e. antibodies or vaccines) humans can rely on. 
In the last months, since the virus was first identified as a novel coronavirus in January 2020, an ever increasing number of scientific works have appeared in literature. Identifying relevant research outcomes at very early stages is utmost important for guiding the scientific community and governments in more effective research and decisions, respectively. However, traditional methods for measuring the relevance and impact of research outcomes (e.g. citation count, impact factor, etc.) might be ineffective due to the extremely narrow observation window currently available. Notoriously, indicators like citation count or impact factor require broader observation windows (i.e. few years) to be reliable~\cite{Lehmann08}. Altmetrics might be valid tools for measuring the impact in quasi-zero-day time-window. Altmetrics\footnote{The altmetrics manifesto: \url{http://altmetrics.org/manifesto/}} have been introduced by~\cite{Priem2012} as the study and use of scholarly impact measures based on activity in online tools and environments. The term has also been used to describe the metrics themselves. COVID-19 pandemic offers an extraordinary playground for understanding inherent correlation between impact and altmetrics. In fact, for the first time in human history, we are facing a pandemic, which is described, debated, and investigated in real time by the scientific community via conventional research venues (i.e. journal papers), jointly with social and on-line media.

In this work we investigate the following research questions:
\begin{itemize}
    \item {\em RQ1}: Which are the platforms, systems and tools mostly used for early scholarly communication?
    \item {\em RQ2}: How is it possible to use altmetrics for automatically identifying candidate impactful research works in quasi-zero-day time-windows?
\end{itemize}

For answering aforementioned research questions we carry out an experiment by using a sample of 212 papers on COVID-19. This sample has been collected by means of a rigorous literature review.

The rest of the paper is organised in the following way: Section~\ref{sec:background} presents related work; Section~\ref{sec:material} describes the material and method used for the experiments; Section~\ref{sec:exp_setup} presents the data analysis we perform and the results we record; Section~\ref{sec:discussion} discusses the results; finally, Section~\ref{sec:conclusion} presents our conclusions and future work.

\section{Related work}
\label{sec:background}
An ever increasing amount of research work has investigated the role of altmetrics in measuring impact since they have been introduced by~\cite{Priem2012}. 

{\bf Correlation among indicators.}
Much research focuses on finding a correlation between altmetrics and traditional indicators. The rationale behind these works is based on the assumption that traditional indicators have been extensively used for scoring research works, and measuring their impact. Hence, their reliability is accepted by practice. Works such as~\cite{Li2012_1,Li2012_2,Bar2012,Thelwall2013,Sud2014,Nuzzolese2019} follow this research line. These studies record moderate agreement (i.e. $\sim$0.6 with Spearman correlation coefficient) with specific sources of altmetrics, i.e. Mendeley and Twitter. According to~\cite{Thelwall2018} and~\cite{Nuzzolese2019}, Mendeley is the on-line platform that provides an indicator (i.e. the number of Mendeley readers) that correlates well with citation counts after a time period consisting of few years. The meta-analyisis conducted by~\cite{Bornmann2015alternative} confirms this result, i.e. the correlation with traditional citations for micro-blogging  is negligible, for blog counts it is small, and for bookmark counts from online reference managers, it is medium to large.
Nevertheless, none of those studies take into account the key property of altmetrics, i.e. they emerge quickly~\cite{Peters2014}. Hence, altmetrics should be used for measuring impact at very early stages, as soon as a topic emerges or a set of research works appear in literature. As a consequence, we use a tight time scale (i.e. quasi-zero-day time-window) for carrying out our analysis.

{\bf Altmetrics and research impact.}
The analysis of altmetrics with respect to research evaluation frameworks has been carried out by~\cite{Wouters2015metric,Ravenscroft2017measuring,Bornmann2018altmetrics,Nuzzolese2019}. More in detail,~\cite{Wouters2015metric} uses the Research Excellence Framework (REF) 2016, i.e. the reference system for assessing the quality of research in UK higher education institutions, for mining possible correlation among different metrics. The analysis is based on different metrics (either traditional or alternative) and research areas, and its outcomes converge towards limited or no correlation.
\cite{Ravenscroft2017measuring} finds very low or negative correlation coefficients between altmetrics provided by Altmetric.com and REF scores concerning societal impact published by British universities in use case studies.  \cite{Bornmann2018altmetrics} investigates the correlation between citation counts and the relationship between the dimensions and quality of papers using regression analysis on post-publication peer-review system of F1000Prime assessments. Such a regression analysis shows that only Mendeley readers and citation counts are significantly related to quality. Finally,~\cite{Nuzzolese2019} uses the data from the Italian National Scientific Qualification (NSQ). The results show good correlation between Mendeley readers and citation count, and moderate accuracy for the automatic prediction of the candidates’ qualification at the NSQ by using independent settings of indicators as features for training a N\"{a}ive Bayes algorithm. 

Some of the aforementioned works focuses on providing a comprehensive analysis investigating not only the correlation between traditional indicators and altmetrics, but also the correlation among the altmetrics themselves. However, all of them overlook the time constraint (i.e. a tight observation window), which is utmost important in our scenario.

\section{Material and method}
\label{sec:material}
In this section we present the input data and the method used for processing such data. More in detail, we explain: (i) the approach adopted for carrying out the literature review focused on gathering relevant literature associated with the COVID-19 pandemic; (ii) the sources and the solution used for enriching the resulting articles with citation count as well as altmetrics; and (iii), finally, the method followed for processing collected data.

\subsection{Literature review}
\label{sec:literature_review}
The initial search was implemented on February 17th, 2020 in MEDLINE/Pub\-med. The search query consists of the following search terms selected by the authors to describe the new pandemic: [coronavirus* OR Pneumonia of Unknown Etiology OR COVID-19 OR nCoV]. Although the name has been updated to SARS-CoV-2 by the International Committee on Taxonomy of Viruses\footnote{\url{https://www.nature.com/articles/s41564-020-0695-z}} on February 11th 2020, the search is performed by using the term ``nCoV'' because we presume that no one, between February 11th and 13th, would have used the term ``SARS-COV-2''. Furthermore, the search is limited to the following time-span: from January 15th, 2019 to February 24th, 2020. Due to the extraordinary rapidity, with which scientific papers have been electronically published online (i.e. ePub), it may happen that some of these have indicated a date later than February 13th 2020 as publication date.

We rely on a two-stage screening process to assess the relevance of studies identified in the search. For the first level of screening, only the title and abstract are reviewed to preclude waste of resources in procuring articles that do not meet the minimum inclusion criteria. Titles and abstracts of studies initially identified are then checked by two independent investigators, and disagreement among reviewers are resolved through a mediator. Disagreement is resolved primarily through discussion and consensus between the researchers. If consensus is not reached, another blind reviewer acts as third arbiter.

Then, we retrieve the full-text for those articles deemed relevant after title and abstract screening. A form developed by the authors is used to record meta-data such as publication date, objective of the study, publication type, study sector, subject matter, and data sources. Results, reported challenges, limitations, conclusions and other information are ignored as they are out of scope with respect to this study.

{\bf Eligibility criteria.} Studies are eligible for inclusion if they broadly include data and information related to COVID-19 and/or SARS-CoV-2. Because of limited resources for translation, articles published in languages other than English are excluded. Papers that describe Coronaviruses that are not SARS-CoV-2 are excluded. There is no restriction regarding publication status.
In summary, the inclusion criteria adopted are: (i) English language; (ii) SARS-CoV 2; (iii) COVID-19; (iv) Pneumonia of Unknown etiology occurred in China between December 2019 and January 2020.
Instead, exclusion criteria are: (i) irrelevant titles not indicating the research topic; (ii) coronavirus not SARS-CoV 2; (iii) SARS, MERS, other coronavirus-related disease not COVID-19; (iv) not human diseases. 

{\bf Data summary and synthesis.} The data are compiled in a single spreadsheet and imported into Microsoft Excel 2010 for validation. Descriptive statistics were calculated to summarize the data. Frequencies and percentages are utilized to describe nominal data. In next section (cf. Section~\ref{sec:data_workflow}) we report statistics about collected papers.

\subsection{Data processing workflow}
\label{sec:data_workflow}
\begin{figure}[!ht]
\centering
\includegraphics[scale=0.075]{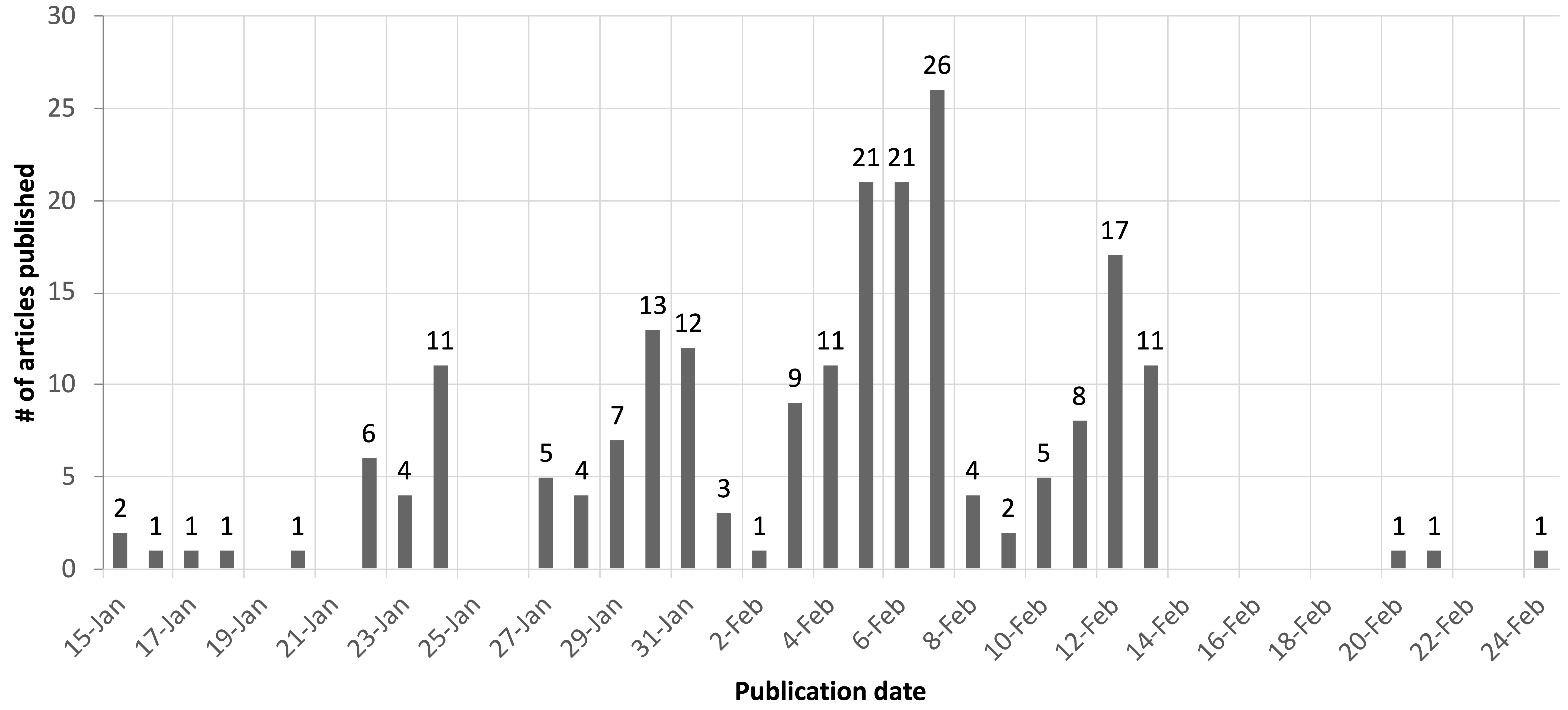}
\caption{Number of papers per publication date in the time-window ranging from January 15th, 2020 to February 24th, 2020.} \label{fig:timeline}
\end{figure}

Selected papers resulting from the literature review are used as input of the data processing workflow. The latter allows us to automatically gather quantitative bibliometric indicators and altmetrics about selected papers and to organise them in a structured format consisting of a knowledge graph. Figure~\ref{fig:timeline} shows the number of papers in the sample grouped by publication date.

\begin{figure}[!ht]
\centering
\includegraphics[scale=0.5]{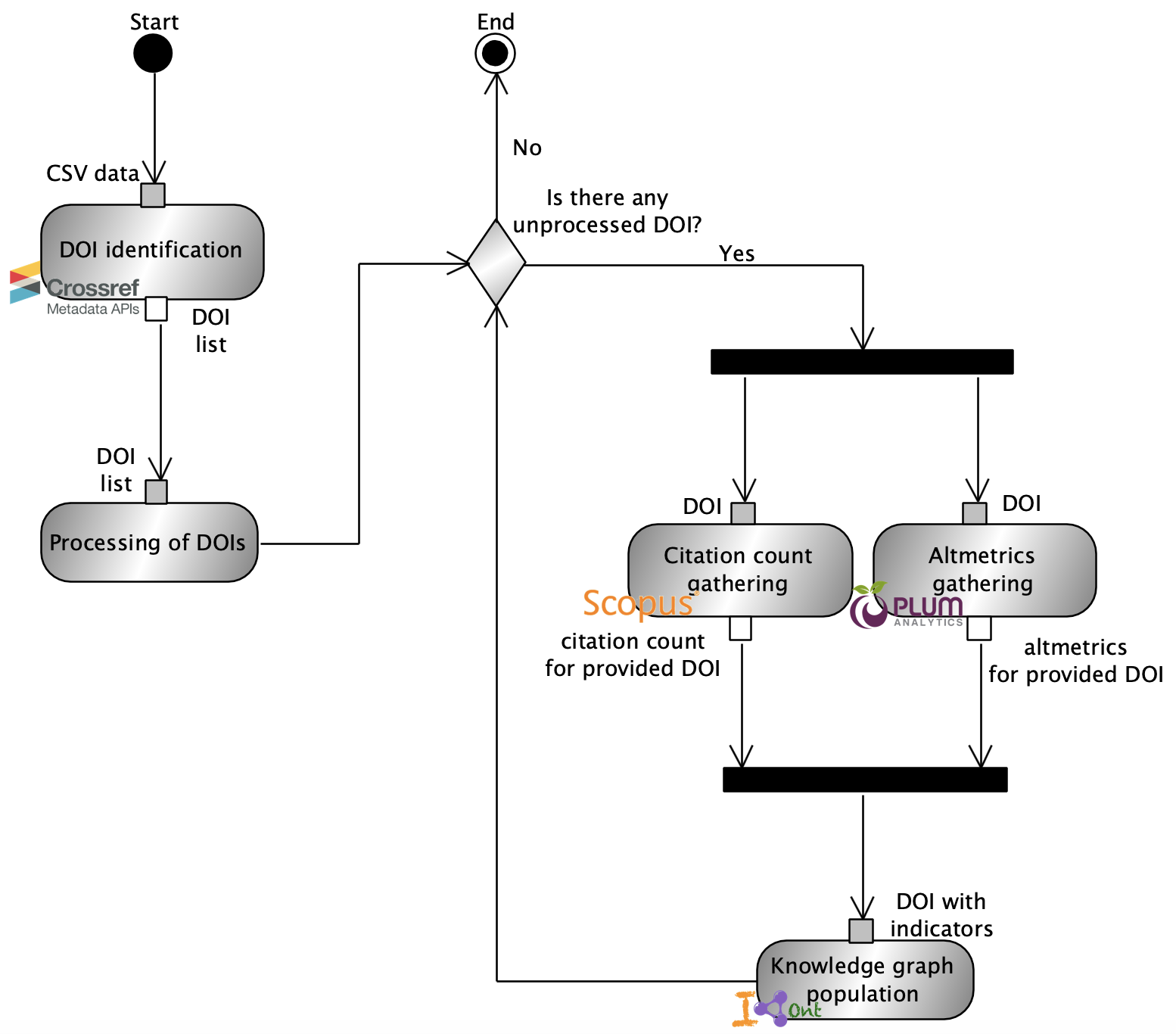}
\caption{The UML activity diagram that graphically represents the workflow re-used and extended from (Nuzzolese et al. 2019) for the data processing activities.} 
\label{fig:workflow}
\end{figure}

The workflow is based on an extension of the one we presented in~\cite{Nuzzolese2019}. Figure~\ref{fig:workflow} shows the workflow as an UML activity diagram. In the diagram: (i) gray rectangles represent activities (e.g. the rectangle labelled ``DOI identification''); (ii) gray boxes represent activities’ input pins; and (iii) white boxes represent activities' output pins. The first activity is the identification of DOIs associated with selected papers. This is performed by processing the spreadsheet resulting from the literature review (cf. Section~\ref{sec:literature_review}). Such a spreadsheet contains an article for each row. In turn, for each row, we take into account the following columns: (i) the internal identifier used for uniquely identifying the article within the CSV, (ii) the authors, (iii) the paper title, and (iv) the DOI whenever possible.

We rely on the Metadata API provided by Crossref\footnote{\url{https://github.com/CrossRef/rest-api-doc}} for checking available DOIs and retrieving missing ones. This API is queried by using the first author and the title associated with each article as input parameters. Crossref returns the DOI that matches the query parameters as output. Whether a DOI is already available we first get the DOI from Crossref, then we check that the two DOIs (i.e. the one already available and one gathered from Crossref) are equal. In case the two DOIs are not equal we keep the DOI gathered from Crossref as valid. This criterion is followed in order to fix possible manual errors (e.g. typos) that would prevent the correct execution of subsequent actions of the workflow. 

The output of the DOI identification activity is a list of DOIs which is passed as input to the second activity named ``Processing of DOIs''. The latter iterates over the list of DOIs and selects them one by one. This operation allows other activities to gather information about citation count and altmetrics by using the DOI as the key for querying dedicated web services. The processing of DOIs proceeds until there is no remaining unprocessed DOI in the list (cf. the decision point labelled ``Is there any unprocessed DOI?'' in Figure~\ref{fig:workflow}).

The activities ``Citation count gathering'' and ``Altmetrics gathering'' are carried out in parallel. Both accept a single DOI as input parameter and return the citation count and the altmetrics associated with such a DOI, respectively. The citation count gathering relies on the API provided by Scopus\footnote{\url{https://dev.elsevier.com/tecdoc_cited_by_in_scopus.html}}. We use Scopus as it is used by many organisations as the reference service for assessing the impact of research from a quantitative perspective (e.g. citation count, $h$-index, and impact factor). For example, the Italian National Scientific Habilitation\footnote{\url{https://www.anvur.it/en/activities/asn/}} (ASN) uses Scopus for defining threshold values about the number of citations and $h$-index scores that candidates to permanent positions of Full and Associate Professor in Italian universities should exceed. The altmetrics gathering activity is based on Plum Analytics\footnote{\url{https://plumanalytics.com/learn/about-metrics/}} (PlumX), which is accessed through its integration in the Scopus API\footnote{\url{https://dev.elsevier.com/documentation/PlumXMetricsAPI.wadl}}. We use PlumX among the variety of altmetrics providers (e.g. Altmetric.com or ImpactStory) as, according to~\cite{Peters2014}, it is the service that registers the most metrics for the most platforms. Additionally, in our previous study~\cite{Nuzzolese2019}, we found that PlumX is currently the service that covers the highest number of research work ($\sim$52.6M\footnote{Data retrieved from \url{https://plumanalytics.com/learn/about-metrics/coverage/} on March 2020.}) if compared to Altmetric.com ($\sim$5M\footnote{Data retrieved from \url{https://figshare.com/articles/Altmetric_the_story_so_far/2812843/1} on March 2020.}) and ImpactStory ($\sim$1M\footnote{Data retrieved from \url{https://twitter.com/Impactstory/status/731258457618157568}  on March 2020.}). PlumX provides three different levels of analytics consisting of (i) the category, which provides a global view across different indicators that are similar in semantics (e.g. the number of alternative citations a research work collects on social media); (ii) the metric, which identifies the indicator (e.g. the number of tweets about a research work); (iii) and the source, that basically allows to track the provenance of an indicator (e.g. the number of tweets on Twitter about a research work). Hereinafter we refer to these levels as the category-metric-source hierarchy. Table~\ref{tab:plumx_info} summarises the categories provided by PlumX by suggesting an explanation for each of them. A more detailed explanation about the categories, metric, and sources as provided by PlumX is available on-line\footnote{\url{https://plumanalytics.com/learn/about-metrics/}}.

\begin{center}
\begin{table}[!ht]
\centering
\caption{The categories  provided by PlumX along with an explanation about their semantics.} 
\label{tab:plumx_info}
\resizebox{0.9\textwidth}{!}{ 
\begin{tabular}{p{1.5cm}|p{3.5cm}||p{1.5cm}|p{3.5cm}}
\centering
 {\bf Category } & {\bf Explanation } & {\bf Category } & {\bf Explanation } \\ \hline \hline
{\bf Usage } & 
A signal that anyone is reading an article or otherwise using a research. & {\bf Captures } & 
An indication that someone wants to come back to the work. \\ \hline
{\bf Mentions } & Number of mentions retrieved in news articles or blog posts about research. & {\bf Social Media } & The number of mentions included in tweets, Facebook likes, etc. that reference a research work. \\
\end{tabular}
}
\end{table}
\end{center}

Once the information about the citation count and altmetrics for an article is available, it is used for populating a knowledge graph in the activity labelled ``Knowledge graph population''. The knowledge graph is represented as RDF and modelled by using the Indicators Ontology (I-Ont)~\cite{Nuzzolese2018}. I-Ont is an ontology for representing scholarly artefacts (e.g. journal articles) and their associated indicators, e.g. citation count or altmetrics such as the number of readers on Mendeley. I-Ont is designed as an OWL\footnote{\url{https://www.w3.org/TR/owl2-overview/}} ontology and was originally meant for representing indicators associated with the articles available on ScholarlyData. ScholarlyData\footnote{\url{https://w3id.org/scholarlydata}}~\cite{Nuzzolese2016} is the reference linked open dataset of the Semantic Web community about papers, people, organisations, and events related to its academic conferences. The resulting knowledge graph, hereinafter referred to as COVID-19-KG, is available on Zenodo\footnote{\url{https://doi.org/10.5281/zenodo.3748694}} for download. Table~\ref{tab:altmetrics_stats} reports the statistics recorded for the metric categories stored into the knowledge graph. We do not report statistics on minimum values as they are meaningless being them 0 for all categories.

\begin{center}
\begin{table}[!ht]
\centering
\caption{The statistics recorded for metric categories including the citation count.} 
\label{tab:altmetrics_stats}
\resizebox{0.9\textwidth}{!}{ 
\begin{tabular}{p{2.5cm}|p{3.5cm}|p{1.1cm}|p{1.1cm}|p{1.1cm}}
\centering
 {\bf Category } & {\bf Metric } & {\bf Max } & {\bf Mean } & {\bf Median } \\ \hline \hline
\multirow{2}{*}[-0.25em]{\bf Citations } & Citation Count & 82 & 1.63 & 0 \\ \cline{2-5}
 & {\bf Total } & {\bf 82 } & {\bf 1.63 } & {\bf 0 } \\ \hline
\multirow{2}{*}[-0.25em]{\bf Captures } & Readers & 161 & 1.69 & 0 \\ \cline{2-5}
 & {\bf Total } & {\bf 161 } & {\bf 1.69 } & {\bf 0 } \\ \hline
\multirow{2}{*}[-2.2em]{\bf Mentions } & Blog Mentions & 22 & 0.95 & 0 \\ \cline{2-5}
& News Mentions & 253 & 8.95 & 0 \\ \cline{2-5}
& Q\&A Site Mentions & 3 & 0.05 & 0 \\ \cline{2-5}
& References & 4 & 0.06 & 0 \\ \cline{2-5}
& {\bf Total } & {\bf 277 } & {\bf 10.0 } & {\bf 0 } \\ \hline
\multirow{2}{*}[-0.25em]{\bf Usage } & Abstract Views & 15 & 0.07 & 0 \\ \cline{2-5}
  & {\bf Total } & {\bf 15 } & {\bf 0.07 } & {\bf 0 } \\ \hline
 \multirow{2}{*}[-0.9em]{\bf Social Media }
 & Shares, Likes \& Comments
 & 33,043 & 582.06 & 0 \\ \cline{2-5}
 & Tweets & 14,409 & 457.36 & 29.5 \\ \cline{2-5}
 & {\bf Total } & {\bf 45,197 } & {\bf 1,250.34 } & {\bf 36.5 } \\ 
 
\end{tabular}
}
\end{table}
\end{center}

Table~\ref{tab:altmetrics_sources} reports the sources of altmetrics we record in COVID-19-KG. Each source is reported with its corresponding category, metric, and number of articles.
\begin{center}
\begin{table}[!ht]
\centering
\caption{Sources of altmetrics with the numbers of the articles associated with those sources.} 
\label{tab:altmetrics_sources}
\resizebox{0.9\textwidth}{!}{ 
\begin{tabular}{l|l|l|r}
\centering
 {\bf Category } & {\bf Metric } & {\bf Metric } & {\bf \# of articles} \\ \hline \hline
Citations & Citation Count & Scopus & 40 \\ \hline
Capture & Readers & Mendeley & 13 \\ \hline
\multirow{2}{*}[-1em]{Mentions } & Blog Mentions & Blog & 43 \\ \cline{2-4}
 & News Mentions & News & 72 \\ \cline{2-4}
 & Q\&A Site Mentions & Stack Exchange & 8 \\ \cline{2-4}
 & References & Wikipedia & 8 \\ \hline
 \multirow{2}{*}[-0.25em]{Social Media} & Shares, Likes \& Comments & Facebook & 67 \\ \cline{2-4}
 & Tweets & Twitter & 157 \\ \hline
 Usage & Abstract Views & Digital Commons & 1 \\
 
\end{tabular}
}
\end{table}
\end{center}

\begin{figure}[!ht]
\begin{subfigure}[t]{0.5\textwidth}
    \centering
    \includegraphics[scale=0.3]{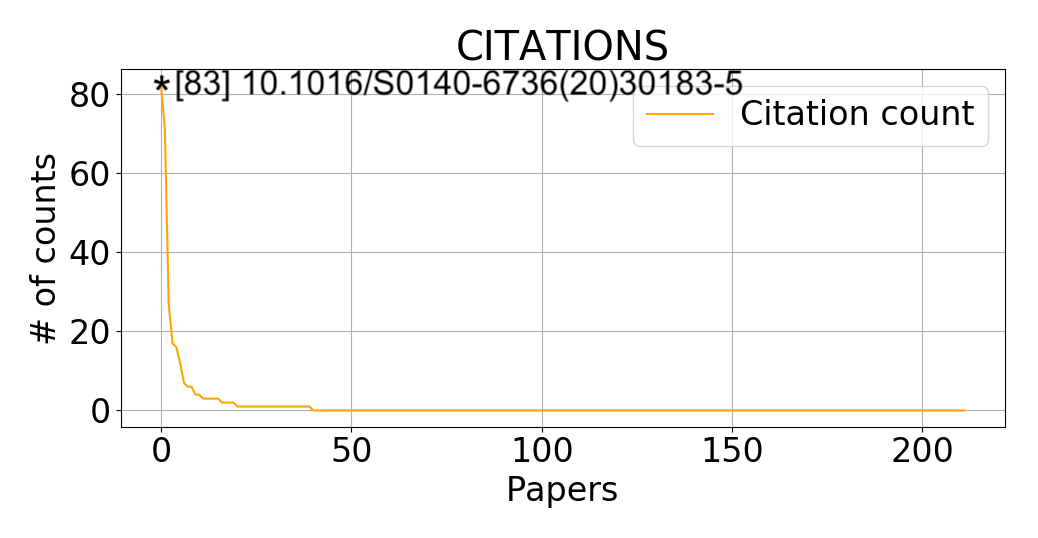}
    \caption{Citations: Citation counts.}
    \label{fig:material_citations}
\end{subfigure}
\begin{subfigure}[t]{0.5\textwidth}
    \centering
    \includegraphics[scale=0.3]{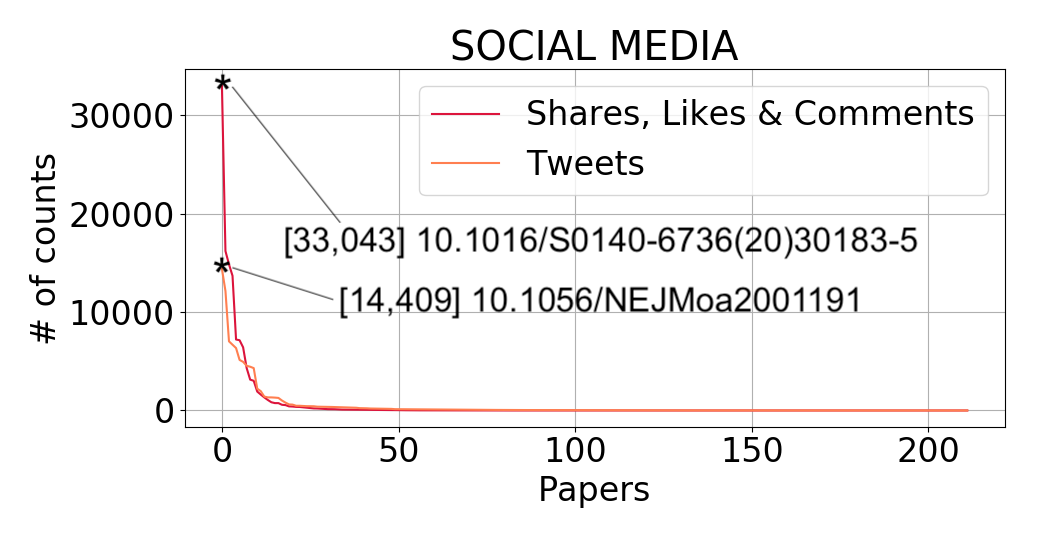}
    \caption{Social Media: Shares, Likes \& Comments, and Tweets.}
    \label{fig:material_social_media}
\end{subfigure}
\begin{subfigure}[t]{0.5\textwidth}
    \centering
    \includegraphics[scale=0.3]{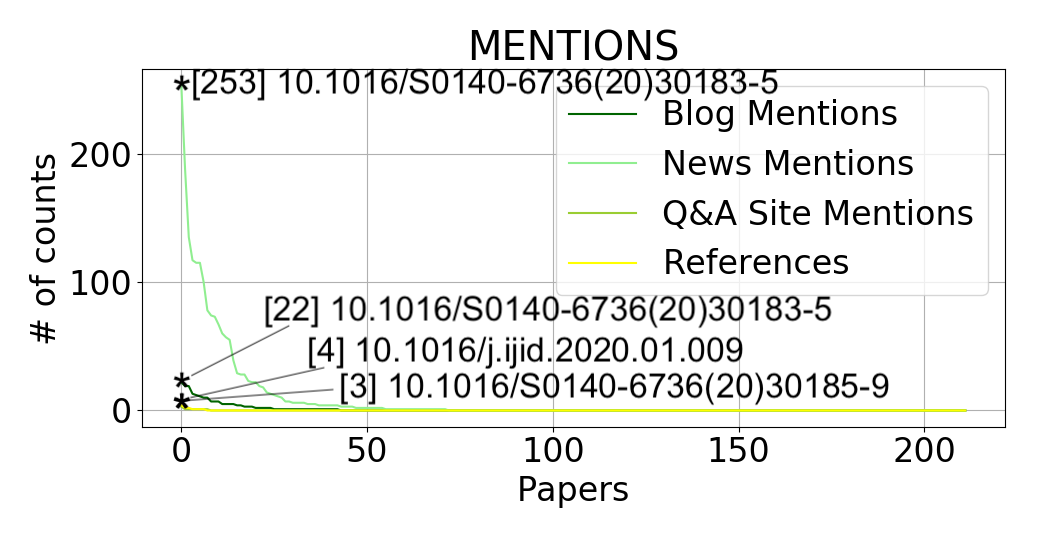}
    \caption{Mentions: Blog, News, Q\&A Site, and References.}
    \label{fig:material_mentions}
\end{subfigure}
\begin{subfigure}[t]{0.5\textwidth}
    \centering
    \includegraphics[scale=0.3]{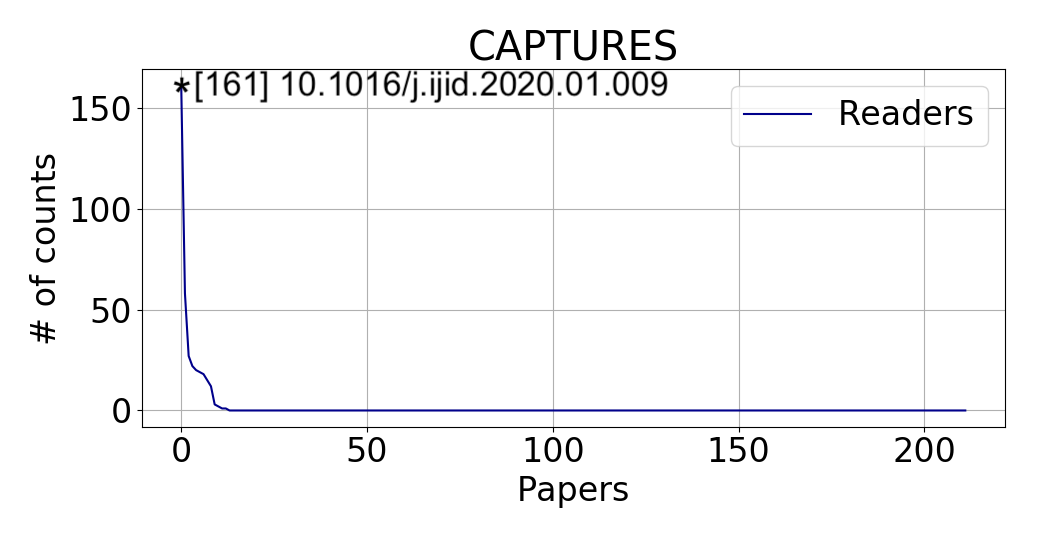}
    \caption{Captures: Readers.}
    \label{fig:material_captures}
\end{subfigure}
\caption{Distributions of the number of indicators per paper.} 
\label{fig:material_categories}
\end{figure}

Figure~\ref{fig:material_categories} shows the distribution of the indicators for each available metric. Namely, Figure~\ref{fig:material_citations} shows the distribution of the citation count over articles; Figure~\ref{fig:material_social_media} shows the distribution of Shares, Likes \& Comments, and Tweets (Social Media category); Figure~\ref{fig:material_mentions} shows the distribution of Blog Mentions, News Mentions, Q\&A Site Mentions, and References (Mentions category); and Figure~\ref{fig:material_captures} shows the distribution of Readers (Captures category) over articles. For each graphic line plotted in Figures~\ref{fig:material_citations}-\subref{fig:material_captures} we report the DOIs associated with the papers that record the highest indicator value for the specific category. That is: (i) the paper with DOI 10.1016/S0140-6736(20)30183-5 appeared in The Lancet records the highest values for citation count (with 83 as value for citation count), ``shares, likes \& comments'' on Facebook (33,043), news mentions (253), and blog mentions (22); (ii) the paper with DOI 10.1016/j.ijid.2020.01.009 appeared in the International Journal of Infectious Diseases records the highest values for Wikipedia references (4) and Mendeley readers (161); (iii) the paper with DOI 10.1056/NEJMoa2001191 appeared in the New England Journal of Medicine records the highest value for tweets on Twitter (14,409); and (iv) the paper with DOI 10.1016/S0140-6736(20)30185-9 appeared in The Lancet records the highest value for ``Q\&A site mentions'' (3).

The workflow is implemented as a Python project and its source code is publicly available on GitHub\footnote{\url{https://github.com/anuzzolese/covid-19-scientometrics}}.

\section{Data analysis}
\label{sec:exp_setup}
We design our experiment in order to address {\em RQ1} and {\em RQ2} by using COVID-19-KG. Hence, we first analyse the different indicators from a behavioural perspective (i.e. {\em RQ1}), i.e. we want to investigate what are the indicators (social media, captures, etc.) and their underlying sources (e.g. Twitter, Mendeley, etc.) that: (i) perform better for scholarly communication in a narrow time-window (i.e. quasi-zero-day); and (ii) share a common intended meaning. Then, we analyse possible methods for identifying candidate impactful research work by relying on available indicators based on the output of the behavioural perspective (i.e. {\em RQ2}).

\subsection{Behavioural perspective}
\label{sec:exp_behaviour}
In order to investigate the behaviour of collected indicators we set up an experiment composed of two conditions: (i) we compute the density estimation for each indicator in the category-metric-source hierarchy first on absolute values, then on standardised values; and (ii) we analyse the correlation among indicators.

{\bf Density estimation.} The density provides a value at any point (i.e. the value associated with an indicator for a given paper) in the sample space (i.e. the whole collection of papers with indicator values in COVID-19-KG). This condition is useful to understand what are possible dense areas for each indicator. The density is computed with the Kernel Density Estimation (KDE)~\cite{Scott2015} by using Gaussian kernels. We use the method introduced by~\cite{Silverman1986} to compute the estimator bandwidth. We remark that the bandwidth is a non-parametric way to estimate the probability density function of a random variable. We opt for~\cite{Silverman1986} as it is one of the most used methods at the state of the art for automatic bandwidth estimation. The KDE is performed first by using absolute values (i.e. the values we record by gathering citation count and altmetrics) as sample set. Then, it is performed by using standardised values as sample set. The former is meant to get the probability distribution for each indicator separately. However, each indicator provides values recorded on very different ranges (cf. Table~\ref{tab:altmetrics_stats}). Hence, KDE resulting from those different indicators are not directly comparable. Accordingly, we standardise indicator values and we then perform KDE over them. Again, KDE is performed for each indicator and for each level of the category-metric-source hierarchy. Standardised values are obtained by computing $z$-scores as the ratio between the sample mean and the standard deviation. Equation~\ref{eq:zeta} formalises the formula we use for computing $z$-scores.

\begin{equation}
\label{eq:zeta}
z(p, i) = \frac{p_{i} - \mu_{i}}{\sigma_{i}}
\end{equation}

In Equation~\ref{eq:zeta}: (i) $p_{i}$ is the value of the indicator $i$ recorded for the paper $p$; (ii) $\mu_{i}$ represents the arithmetic mean computed over the set of all values available for the indicator $i$ for all papers; and (iii) $\sigma_{i}$ represents the standard deviation computed over the set of all values available for the indicator $i$ for all papers.

\begin{figure}[!ht]
\begin{subfigure}[t]{0.5\textwidth}
    \centering
    \includegraphics[scale=0.4]{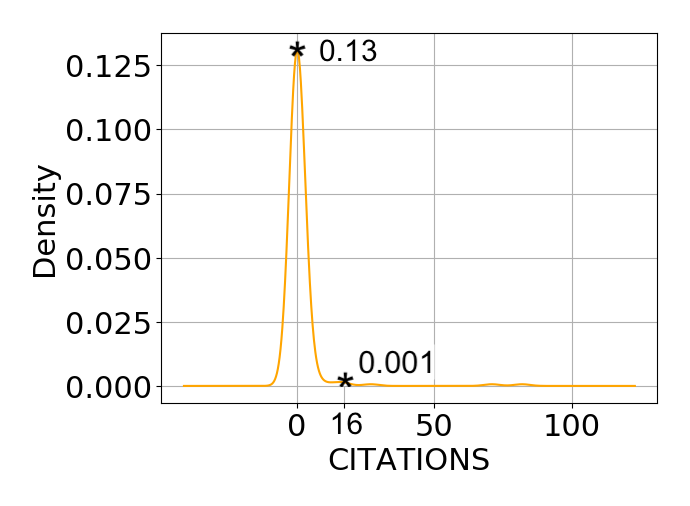}
    \caption{Citations.}
    \label{fig:density_citations}
\end{subfigure}
\begin{subfigure}[t]{0.5\textwidth}
    \centering
    \includegraphics[scale=0.4]{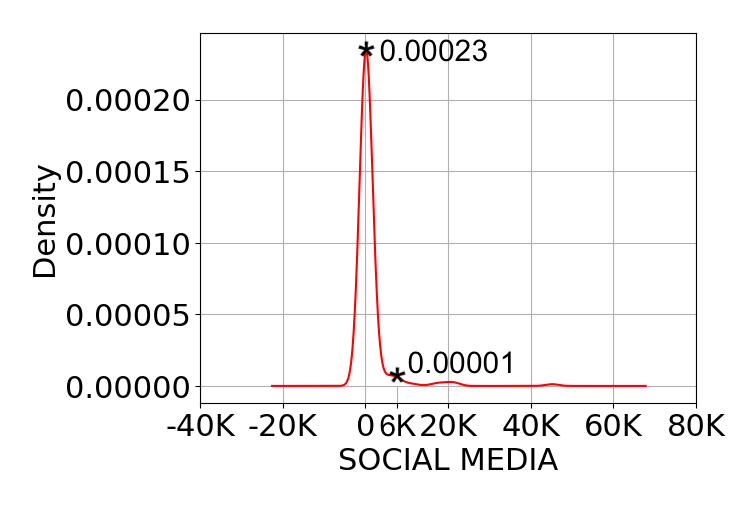}
    \caption{Social Media.}
    \label{fig:density_social_media}
\end{subfigure}
\begin{subfigure}[t]{0.5\textwidth}
    \centering
    \includegraphics[scale=0.4]{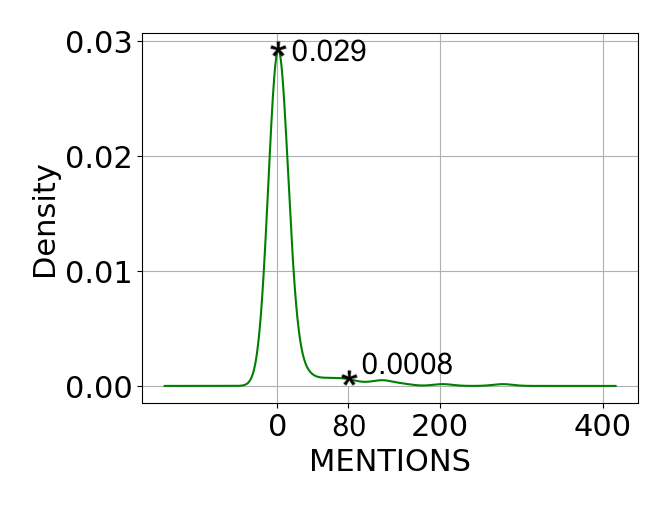}
    \caption{Mentions.}
    \label{fig:density_mentions}
\end{subfigure}
\begin{subfigure}[t]{0.5\textwidth}
    \centering
    \includegraphics[scale=0.4]{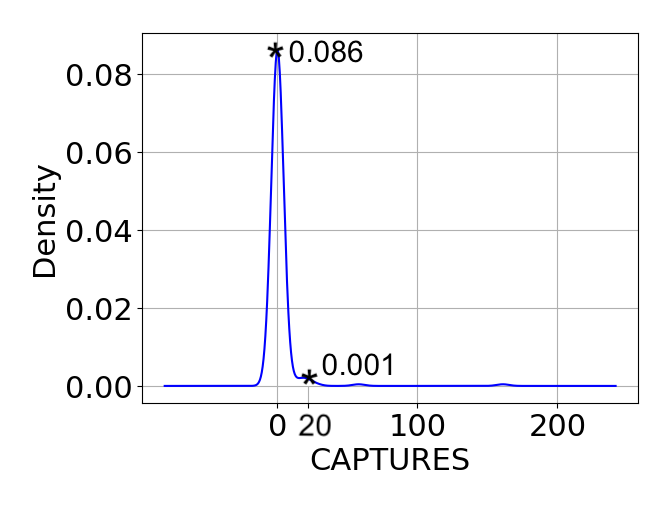}
    \caption{Captures.}
    \label{fig:density_captures}
\end{subfigure}
\caption{Diagrams of the kernel density estimations obtained for each category.} 
\label{fig:density}
\end{figure}

Figure~\ref{fig:density} shows the diagrams of the KDEs we record for each category. For citation counts (cf. Figure~\ref{fig:density_citations}) the most dense area has $d$ ranging from ${\sim}0.13$ and ${\sim}0.001$ and comprises articles that have from 0 to ${\sim}16$ traditional citations. For social media (cf. Figure~\ref{fig:density_social_media}) the most dense area has $d$ ranging from ${\sim}0.00023$ and ${\sim}0.00001$ and comprises articles that have from 0 to ${\sim}6,000$ alternative citations on social media. For mentions (cf. Figure~\ref{fig:density_mentions}) the most dense area has $d$ ranging from ${\sim}0.029$ and ${\sim}0.0008$ and comprises articles that have from 0 to ${\sim}80$ mentions. For captures (cf. Figure~\ref{fig:density_captures}) we record as the most dense area the one having density $d$ ranging from ${\sim}0.08$ and ${\sim}0.001$ and comprising articles that count from 0 to ${\sim}20$ number of captures. We do not compute the KDE for the usage category as there is one article only in COVID-19-KG with a value for such an indicator. 

\begin{figure}[!ht]
\begin{subfigure}[t]{0.5\textwidth}
    \centering
    \includegraphics[scale=0.4]{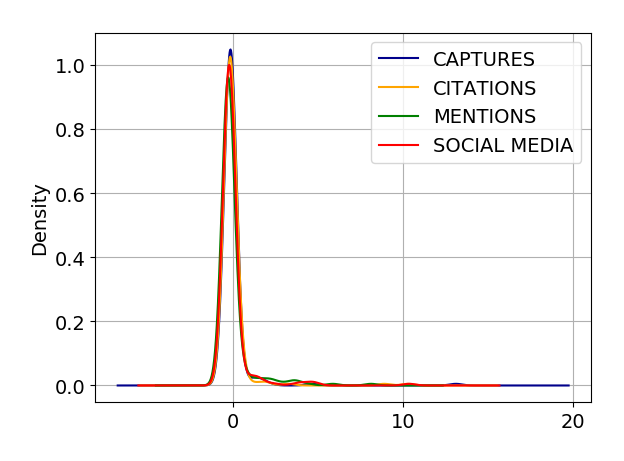}
    \caption{Categories.}
    \label{fig:density_categories}
\end{subfigure}
\begin{subfigure}[t]{0.5\textwidth}
    \centering
    \includegraphics[scale=0.4]{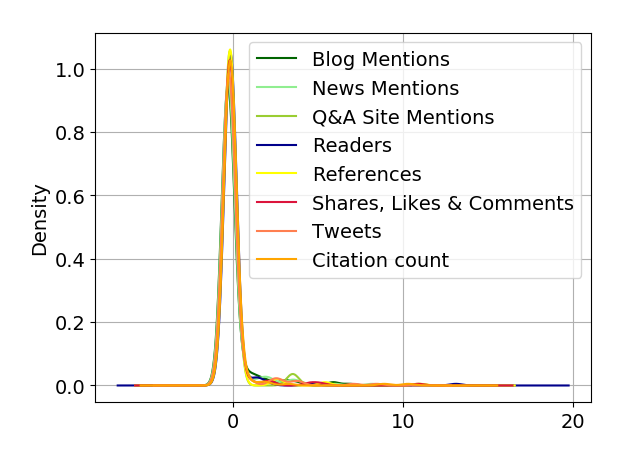}
    \caption{Metrics.}
    \label{fig:density_sources}
\end{subfigure}
\caption{Diagrams of the KDEs computed on $z$-scores for categories (\subref{fig:density_categories}) and metrics (\subref{fig:density_sources}).} 
\label{fig:density_normalised}
\end{figure}

Instead, Figure~\ref{fig:density_normalised} shows the KDE diagrams obtained with the standardised values. More in detail, Figure~\ref{fig:density_categories} and Figure~\ref{fig:density_sources} compare density estimation curves resulting from for the different categories and metrics, respectively. We do not report KDE curves recorded for sources as they are identical to those recorded for metrics. This is due to the fact that there is a one-to-one correspondence between metrics and sources in COVID-19-KG, e.g. the Tweets metric has Twitter only among its sources. All most dense areas are those under the curve determined by $d$ between ${\sim}1$ and ${\sim}0.02$ with values ranging from 0 to ${\sim} 1$ for selected indicators. This is recorded regardless of the specific level of the the category-metric-source hierarchy.

{\bf Correlation analysis.} The correlation analysis aims at identifying similarities among different indicators both in their semantics and intended use on web platforms or social media  (e.g. Twitter, Mendeley, etc.). This investigation is helpful for detecting which indicators are meaningful if used together as tools for computing impact for the articles in our sample. Hence, this analysis, besides being crucial from the behavioural perspective, is preparatory for understanding how to address {\em RQ2}.
This analysis repeats the experiment we carried out in~\cite{Nuzzolese2019}. We remind that in \cite{Nuzzolese2019} we used the papers extracted from the curricula of the candidates to the scientific habilitation process held in Italy for all possible disciplines as dataset. In the context of this work we narrow the experiment to a dataset with very peculiar boundaries in terms of (i) the topic (i.e. COVID-19) and (ii) the observation time-window (i.e. ranging from January 15th 2020 to February 24th 2020). As in \cite{Nuzzolese2019}, we use the sample Pearson correlation coefficient (i.e. $r$) as the measure to assess the linear correlation between pairs of sets of indicators. The Pearson correlation coefficient is widely used in literature. It records correlation in terms of a value ranging from +1 to -1, where, +1 indicates total positive linear correlation, 0 indicates no linear correlation, and -1 indicates total negative linear correlation. For computing $r$, we construct a vector for each paper. The elements of a vector are the indicator values associated with its corresponding paper. We fill elements with 0 if an indicator is not available for a certain paper. The latter condition is mandatory in order to have vectors of equal size. In fact, $r$ is computed by means of pairwise comparisons among vectors. The sample Pearson correlation coefficient is first computed among categories and then on sources by following the category-metric-source hierarchy as provided by PlumX. Again, we do not take into account the level of metrics as it is mirrored by the level of sources with a one-to-one correspondence. Additionally, $r$ is investigated further only for those sources belonging to a category for which we record moderate correlation,  i.e. $r{>} 0.6$. That is, we do not further investigate $r$ if there is limited or no correlation at category level. 

\begin{figure}[!ht]
\begin{subfigure}[t]{0.5\textwidth}
    \centering
    \includegraphics[scale=0.3]{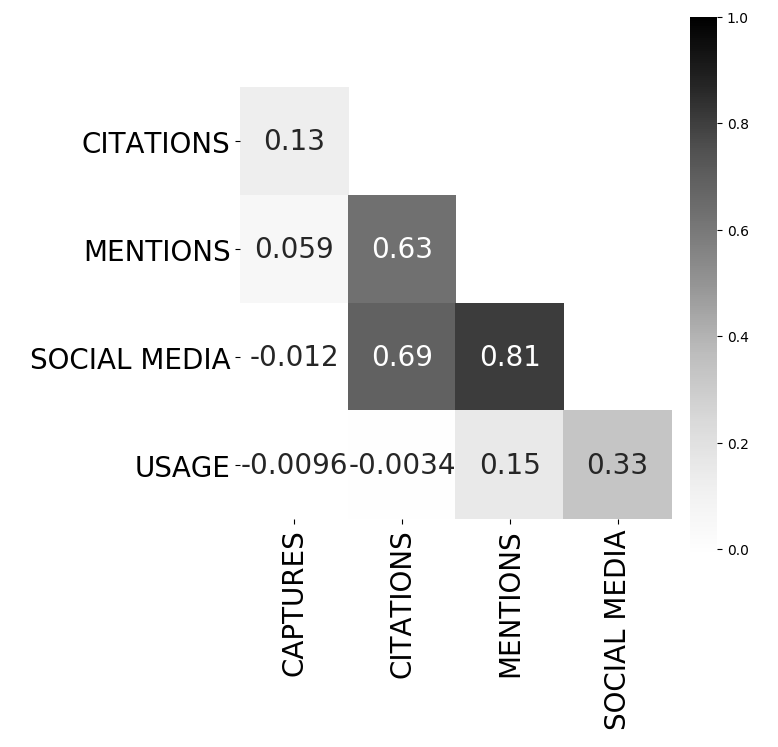}
    \caption{Correlation among categories of indicators.}
    \label{fig:correlation_categories}
\end{subfigure}
\begin{subfigure}[t]{0.5\textwidth}
    \centering
    \includegraphics[scale=0.3]{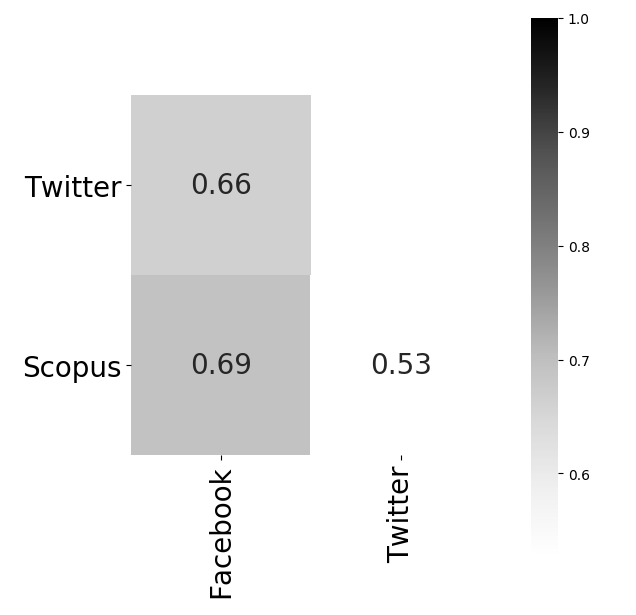}
    \caption{Correlation among Twitter, Facebook, and Scopus citation count.}
    \label{fig:correlation_citations_social_media}
\end{subfigure}
\begin{subfigure}[t]{0.5\textwidth}
    \centering
    \includegraphics[scale=0.3]{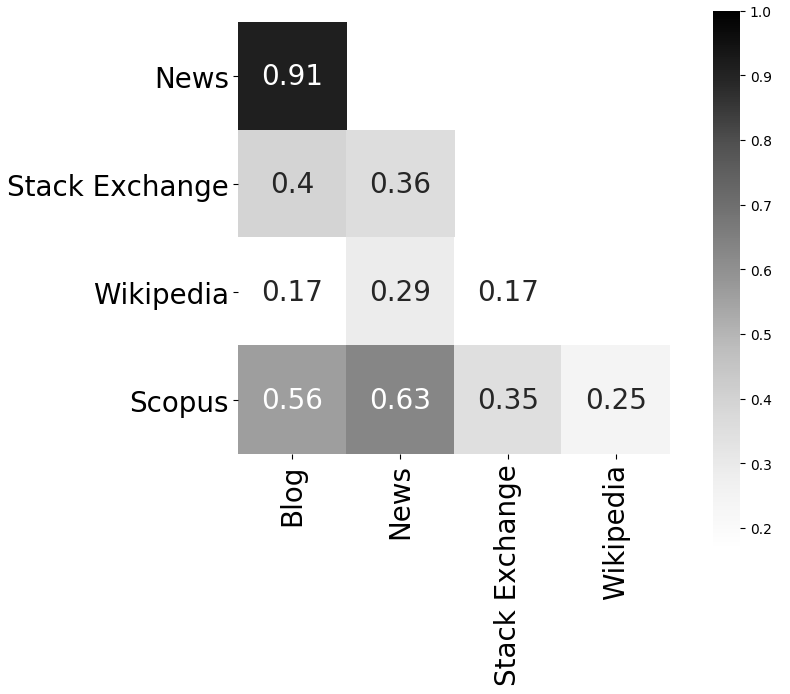}
    \caption{Correlation among News, Stack Exchange, Wikipedia, and Scopus citation count.}
    \label{fig:correlation_citations_mentions}
\end{subfigure}
\begin{subfigure}[t]{0.5\textwidth}
    \centering
    \includegraphics[scale=0.3]{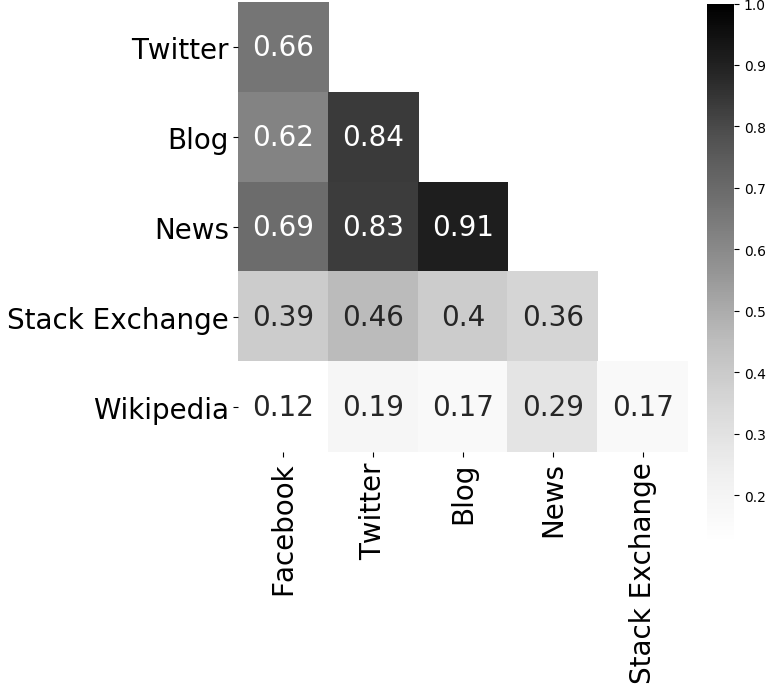}
    \caption{Correlation among News, Stack Exchange, Wikipedia, Twitter, and Facebook.}
    \label{fig:correlation_social_media_mentions}
\end{subfigure}
\caption{Correlation coefficients among indicators.} 
\label{fig:correlation}
\end{figure}

Figure~\ref{fig:correlation} shows the confusion matrices resulting from the pairwise comparisons of the correlation coefficients. For categories (cf. Figure~\ref{fig:correlation_categories}) the highest correlation coefficients are recorded between: (i) mentions and citations, with $r{=}0.63$, statistical significance $p{<}0.01$ ($p$-values are computed by using the Student's $t$-distribution), and standard error $SE_{r}{=\pm{0.04}}$; (ii) social media and citations, with $r{=}0.69$, $p{<}0.01$, and $SE_{r}{=\pm{0.04}}$; and (iii) social media and mentions, with $r{=}0.81$, $p{<}0.01$, and $SE_{r}{=\pm{0.03}}$. Figure~\ref{fig:correlation_citations_social_media} shows the confusion matrix for the sources associated with the social media and citations categories, i.e. Twitter and Facebook for social media and Scopus for citations. If we focus on cross-category sources only (i.e. we do not take into account moderate correlation coefficients recorded between sources associated with the same category) we record moderate correlation between Facebook and Scopus, with $r{=}0.69$, $p{<}0.01$, and $SE_{r}{=\pm{0.04}}$. Figure~\ref{fig:correlation_citations_mentions} shows the confusion matrix for the sources associated with the mentions and citations categories, i.e. News, Stack Exchange, and Wikipedia  for mentions and Scopus for citations. The only cross-category sources associated with moderate correlation are News for mentions and Scopus for citations, with $r{=}0.63$, $p{<}0.01$, and $SE_{r}{=\pm{0.04}}$. Finally, Figure~\ref{fig:correlation_social_media_mentions} shows the confusion matrix for the sources associated with the mentions and social media categories. In the latter we record $r{>}0.6$ for the following cross-category sources: (i) Facebook and News, with $r{=}0.69$, $p{<}0.01$, and $SE_{r}{=\pm{0.04}}$; (ii) Facebook and Blog, with $r{=}0.62$, $p{<}0.01$, and $SE_{r}{=\pm{0.04}}$; (iii) Twitter and News, with $r{=}0.83$, $p{<}0.01$, and $SE_{r}{=\pm{0.03}}$; and (iv) Twitter and Blog, with $r{=}0.84$, $p{<}0.01$, and $SE_{r}{=\pm{0.03}}$.

The strong correlation coefficients recorded among mentions, social media, and citations is in line with the results reported by~\cite{Kousha2020} based on a COVID-19 dataset. However, it is worth noticing that this might be inflated by the large number of zero values recorded as indicators in our sample of articles (cf. Figure~\ref{fig:material_categories}). For further investigating the role played by zero values for computing correlation coefficients we construct four subgraphs from COVID-19-KG. Those subgraphs take into account only the categories that record the highest correlation coefficients (cf. Figure~\ref{fig:correlation_categories}), namely: (i)
$KG_{m,s}$ about articles with only mentions and social media as indicators; (ii) $KG_{m,c}$ about articles with mentions and citations; (iii) $KG_{s,c}$ about articles with social media and citations; (iv) and $KG_{m,s,c}$ about articles with mentions, social media and citations. For each of the aforementioned subgraphs  we filter out the articles having zero values for any of the possible indicators, i.e. all the indicators associated with an article have to be non-negative integers. Accordingly, the four subgraphs contains a different number of articles, that is: (i) 68 for $KG_{m,s}$; (ii) 25 for $KG_{m,c}$; (iii) 39 for $KG_{s,c}$; and, (iv) 25 for $KG_{m,s,c}$, respectively. 


\begin{figure}[t!]
    \begin{subfigure}[t]{0.5\textwidth}
        \centering
        \includegraphics[scale=0.3]{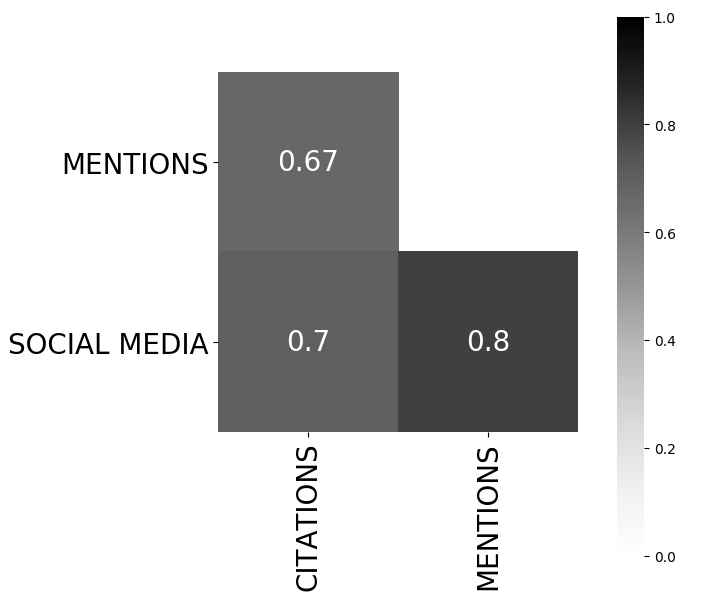}
        \caption{Correlation coefficients recorded for $KG_{m,s}$, $KG_{m,c}$, and $KG_{s,c}$.}
        \label{fig:correlation_categories_no_zeros_separately}
    \end{subfigure}
    \begin{subfigure}[t]{0.5\textwidth}
        \centering
        \includegraphics[scale=0.3]{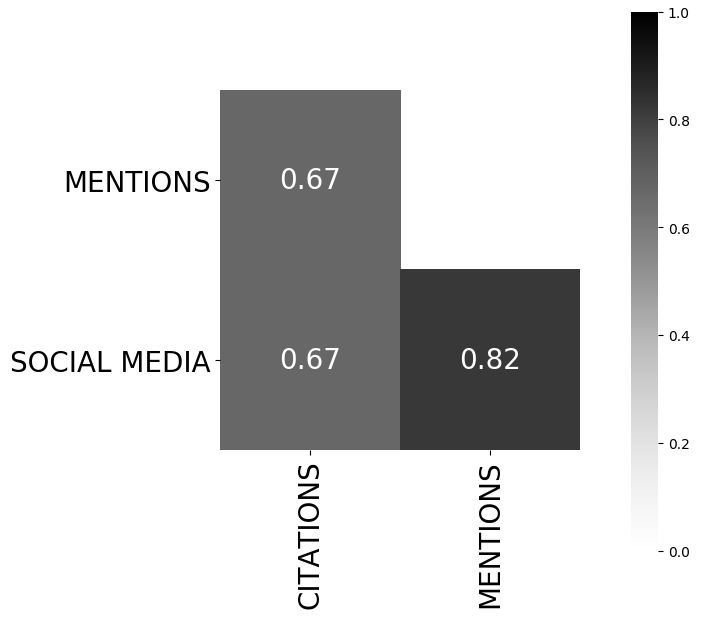}
        \caption{Correlation coefficients recorded for $KG_{m,s,c}$.}
        \label{fig:correlation_categories_no_zeros_all}
    \end{subfigure}
    \caption{Correlation among mentions, social media, and citations with articles without zero-values as indicators.} 
\label{fig:cis}
\end{figure}

\begin{figure}[ht!]
    \centering
    \includegraphics[scale=0.08]{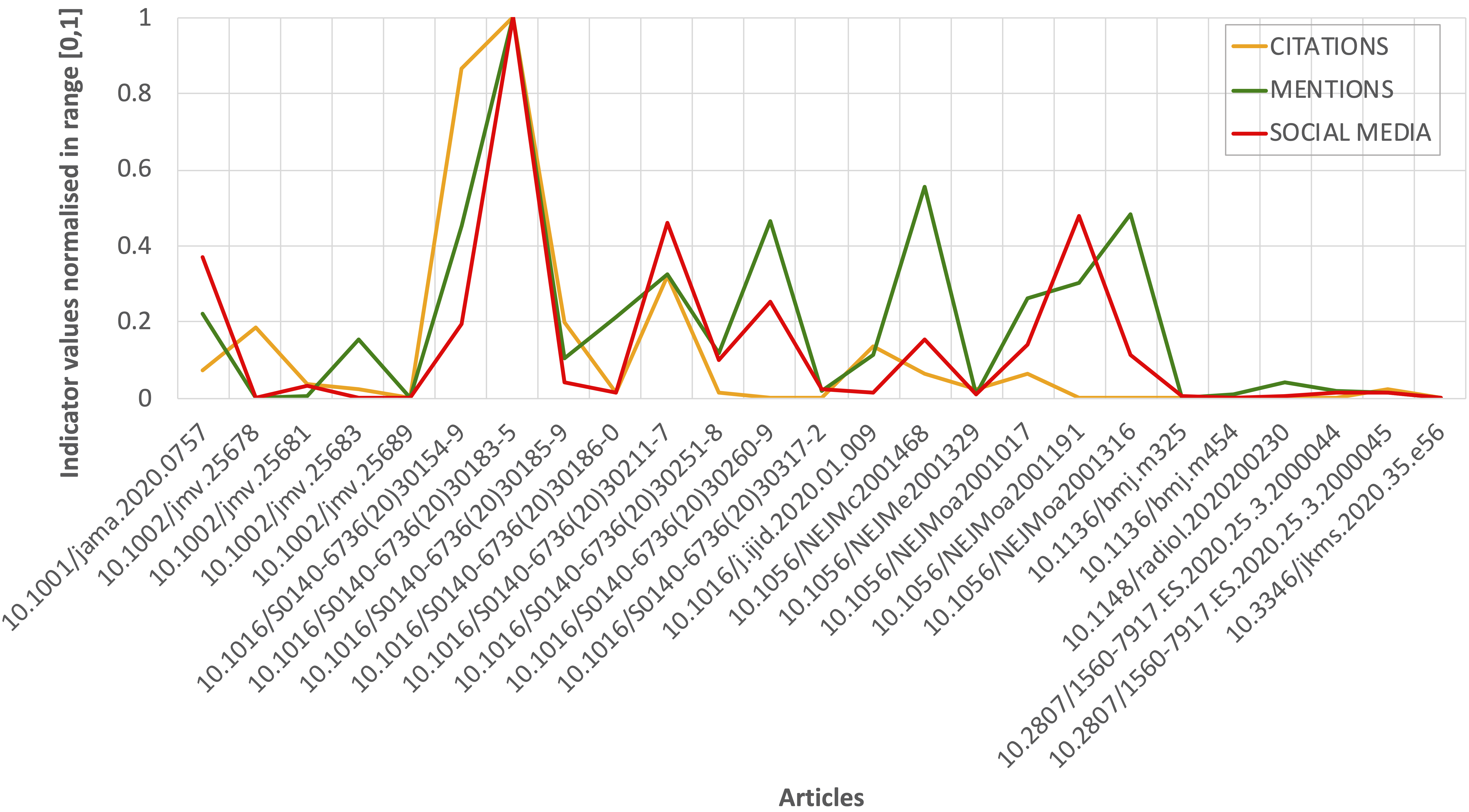}
    \caption{Correlation among categories of indicators with articles with zero-value indicators filtered out from the sample. The values of the indicators for the three categories have been normalised in the range $[0-1]$ for comparison.}
    \label{fig:no_zeros_trends}
\end{figure}

Figure~\ref{fig:correlation_categories_no_zeros_separately} shows the aggregation of the correlation coefficients computed among the pairs of indicators available separately in $KG_{m,s}$, $KG_{m,c}$, and $KG_{s,c}$. We record: (i) moderate correlation between mentions and citations ($r{=}0.67$, statistical significance $p{<}0.01$, and standard error $SE_{r}{=\pm{0.05}}$) and citations and social media ($r{=}0.7$, $p{<}0.01$, and $SE_{r}{=\pm{0.0002}}$); and (ii) strong correlation between social media and mentions ($r{=}0.8$, $p{<}0.01$, and $SE_{r}{=\pm{0.0002}})$. 
Similarly, Figure~\ref{fig:correlation_categories_no_zeros_separately} shows the correlation among mentions, social media and citations as represented in $KG_{m,s,c}$. Again, we record: (i) moderate correlation between mentions and citations ($r{=}0.67$, statistical significance $p{<}0.01$, and standard error $SE_{r}{=\pm{0.05}}$) and citations and social media ($r{=}0.67$, $p{<}0.01$, and $SE_{r}{=\pm{0.0003}}$); and (ii) strong correlation between social media and mentions ($r{=}0.82$, $p{<}0.01$, and $SE_{r}{=\pm{0.0007}})$. In Figure~\ref{fig:no_zeros_trends} it is fairly evident how the values for the indicator categories of mentions, social media, and citations behave similarly. Those values are obtained by normalising the absolute values in the range $[0,1]$ (cf. y-axis) for the 25 articles available into $KG_{m,s,c}$. 

\subsection{Selecting impactful articles}
We then investigate how indicators can be used for selecting candidate qualitative articles among those available in COVID-19-KG. For this analysis we exploit the result of the behavioural perspective. More specifically, we use the outcomes provided by the correlation analysis for defining a strategy in the selection of indicators in our methodology.

\begin{figure}[!ht]
\begin{subfigure}[t]{\textwidth}
    \centering
    \includegraphics[scale=0.28]{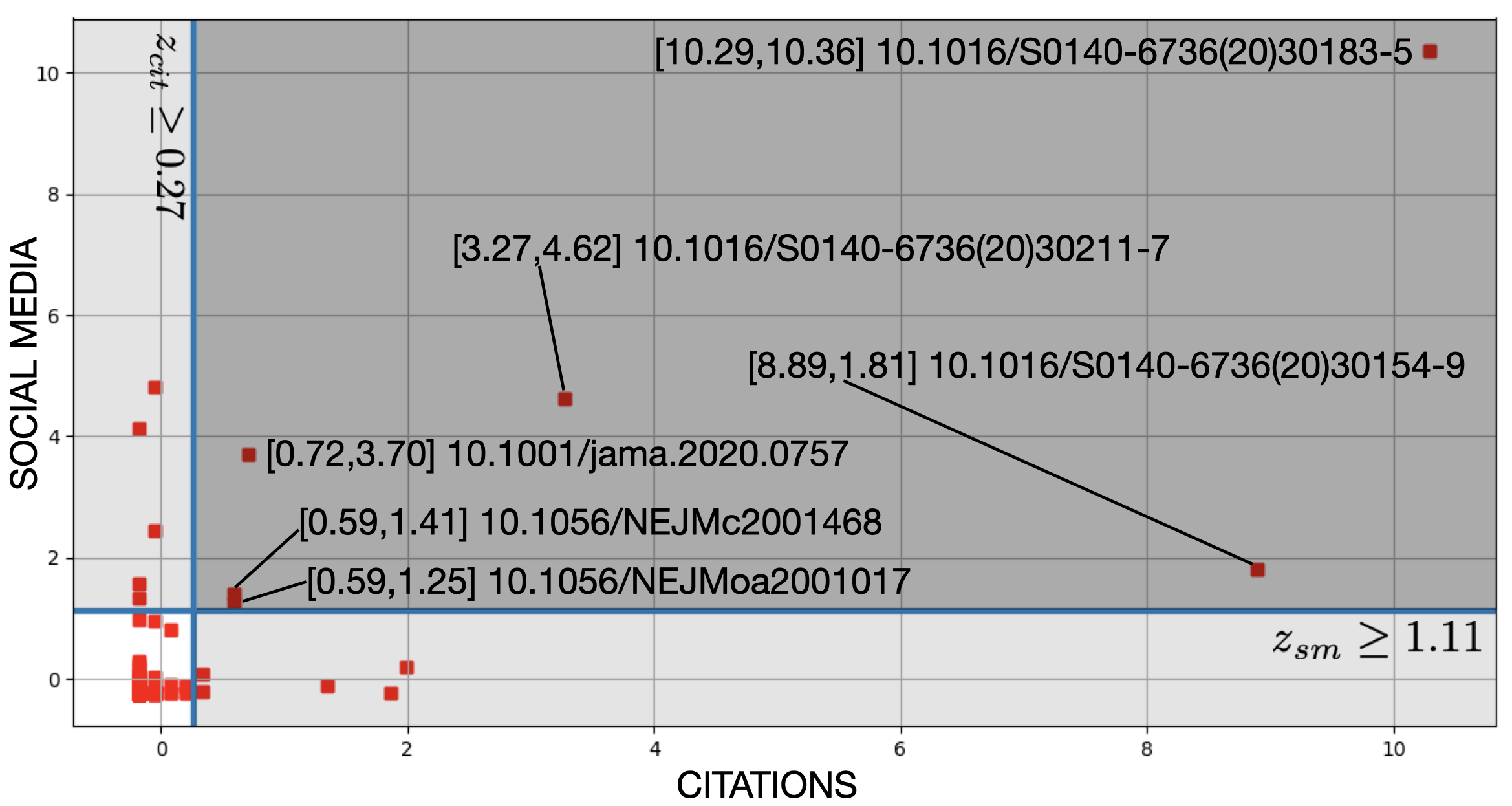}
    \caption{Citations and social media as axis.}
    \label{fig:ratio_citations_social_media}
\end{subfigure}
\begin{subfigure}[t]{\textwidth}
    \centering
    \includegraphics[scale=0.28]{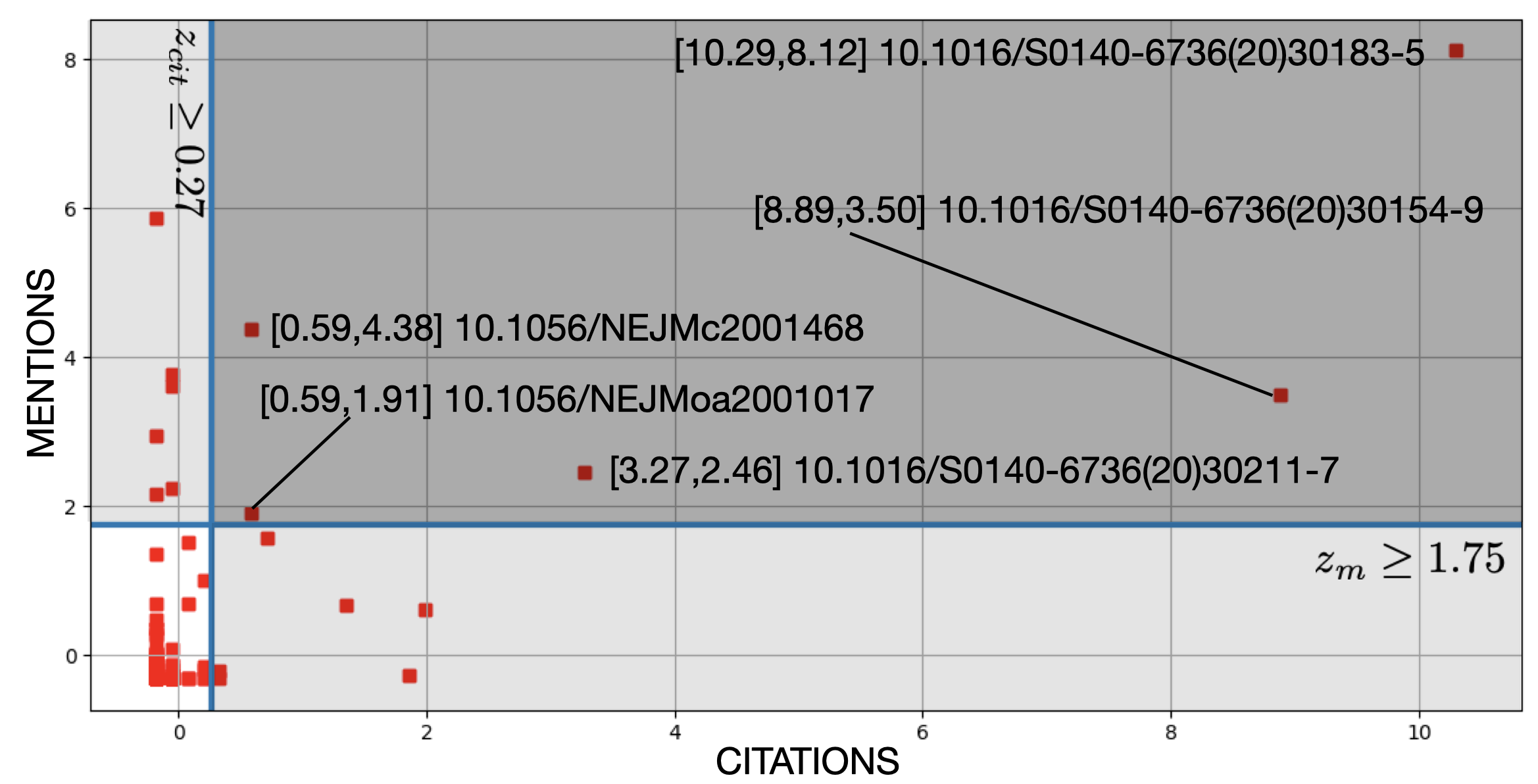}
    \caption{Citations and mentions as axis.}
    \label{fig:ratio_citations_mentions}
\end{subfigure}
\begin{subfigure}[t]{\textwidth}
    \centering
    \includegraphics[scale=0.28]{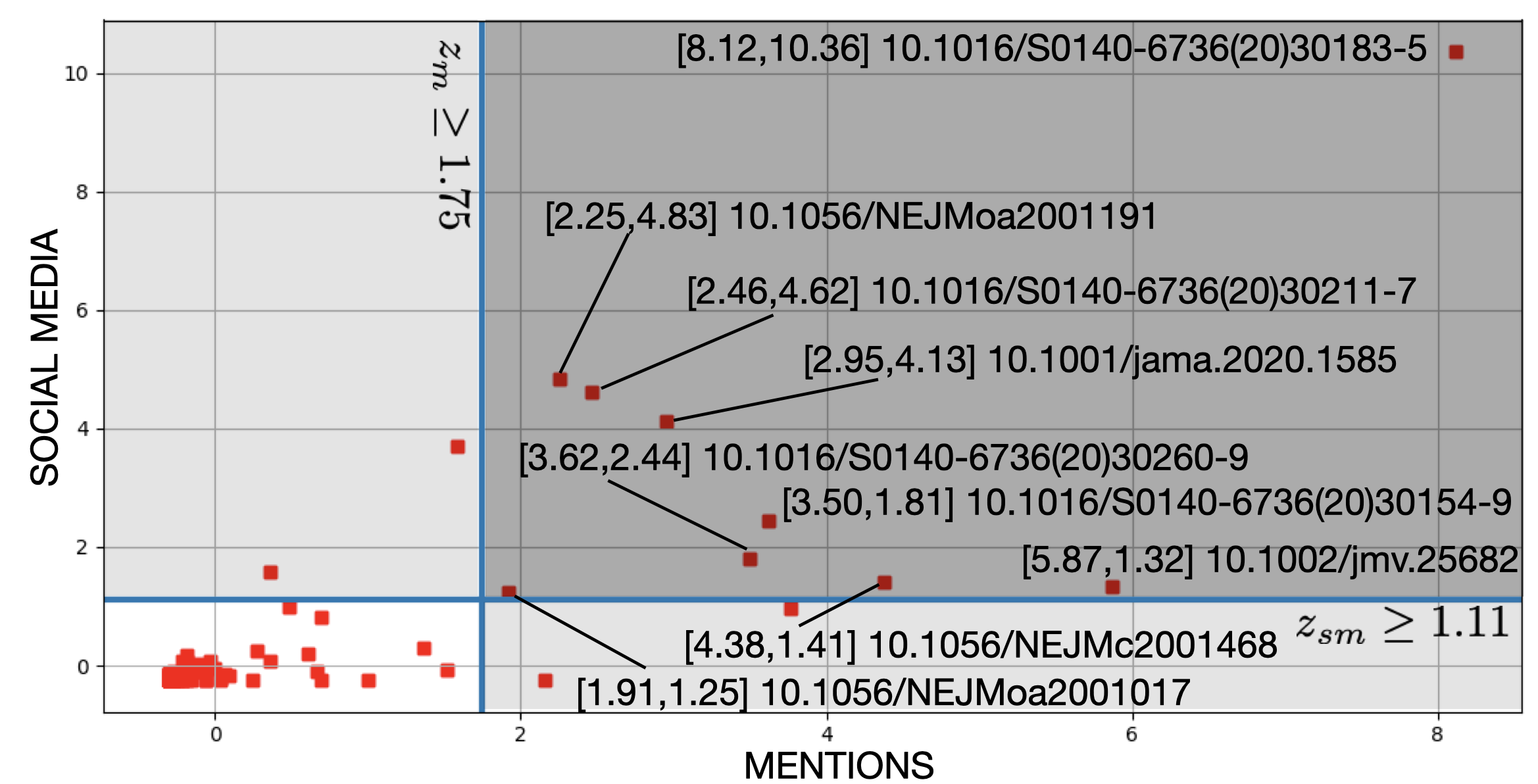}
    \caption{Social media and mentions as axis.}
    \label{fig:ratio_mentions_social_media}
\end{subfigure}
\caption{Geometric spaces.} 
\label{fig:ratio}
\end{figure}

{\bf Geometric selection.} We use the pairs of indicators with moderate to strong correlation coefficients for positioning papers on a Cartesian plane. Then we use such a positioning for defining a selection criterion. The axes of the Cartesian plane are the two indicators part of a pair. The axes values are the $z$-scores computed for each indicator (cf. Equation~\ref{eq:zeta}). We perform this analysis for the pairs (citations, social media), (citations, mentions), and (social media, mentions). Again, we select these pairs only as they correlate better than others according to the correlation analysis (cf. Section~\ref{sec:exp_behaviour}). Furthermore, in COVID-19-KG citations, social media, and mentions are available for the most papers (cf. Figure~\ref{tab:altmetrics_sources}).
Figure~\ref{fig:ratio} shows the results of this analysis. In order to draw a boundary around candidate impactful papers, we identify a threshold $t$ for each category of indicators. We use the lower bound of the 95\% quantiles, i.e. $Q_{95}$, as $t$. The quantiles are obtained by dividing the indicator values available for a given category (e.g. social media) COVID-19-KG into subsets of equal sizes. The lower bounds of the 95\% quantiles recorded are 0.27, 1.11, and 1.75 for citations, social media, and mentions, respectively. For example, the $Q_{95}$ for the citations category contains all that papers that have a $z$-score greater than or equal to 0.27. We opt for 95\% quantiles as they are selective. In fact, they allow us to gather the 5\% papers in COVID-19-KG that record the highest value with respect to the selected indicator categories. 
When we use citations and social media categories (we refer this combination to as $G_{c,s}$) as the axis of the Cartesian plane we record 6 papers whose indicator values are in $Q_{95}$ of both categories (cf. Figure~\ref{fig:ratio_citations_social_media}). Instead, when we use citations and mentions categories (i.e. $G_{c,m}$) as axis we record 5 papers whose indicator values are in $Q_{95}$ of both categories (cf. Figure~\ref{fig:ratio_citations_mentions}). Finally, when we use social media and mentions categories (i.e. $G_{s,m}$) as axis we record 9 papers whose indicator values are in $Q_{95}$ of both categories (cf. Figure~\ref{fig:correlation_social_media_mentions}).

\begin{figure}[!ht]
\begin{subfigure}[t]{\textwidth}
    \centering
    \includegraphics[scale=0.33]{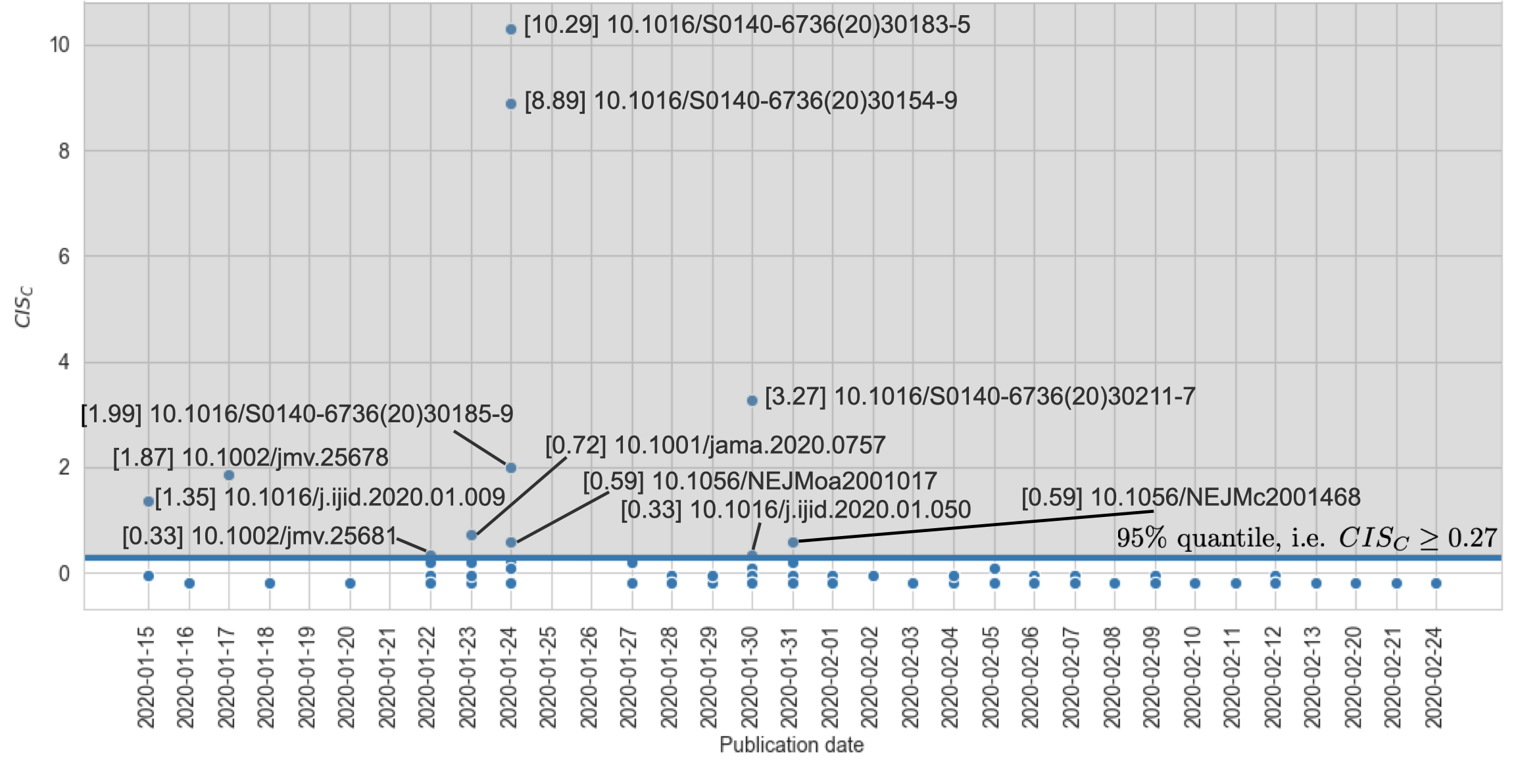}
    \caption{CIS computed on the set indicators $C$ limited to citations.}
    \label{fig:cis_citations}
\end{subfigure}
\begin{subfigure}[t]{\textwidth}
    \centering
    \includegraphics[scale=0.33]{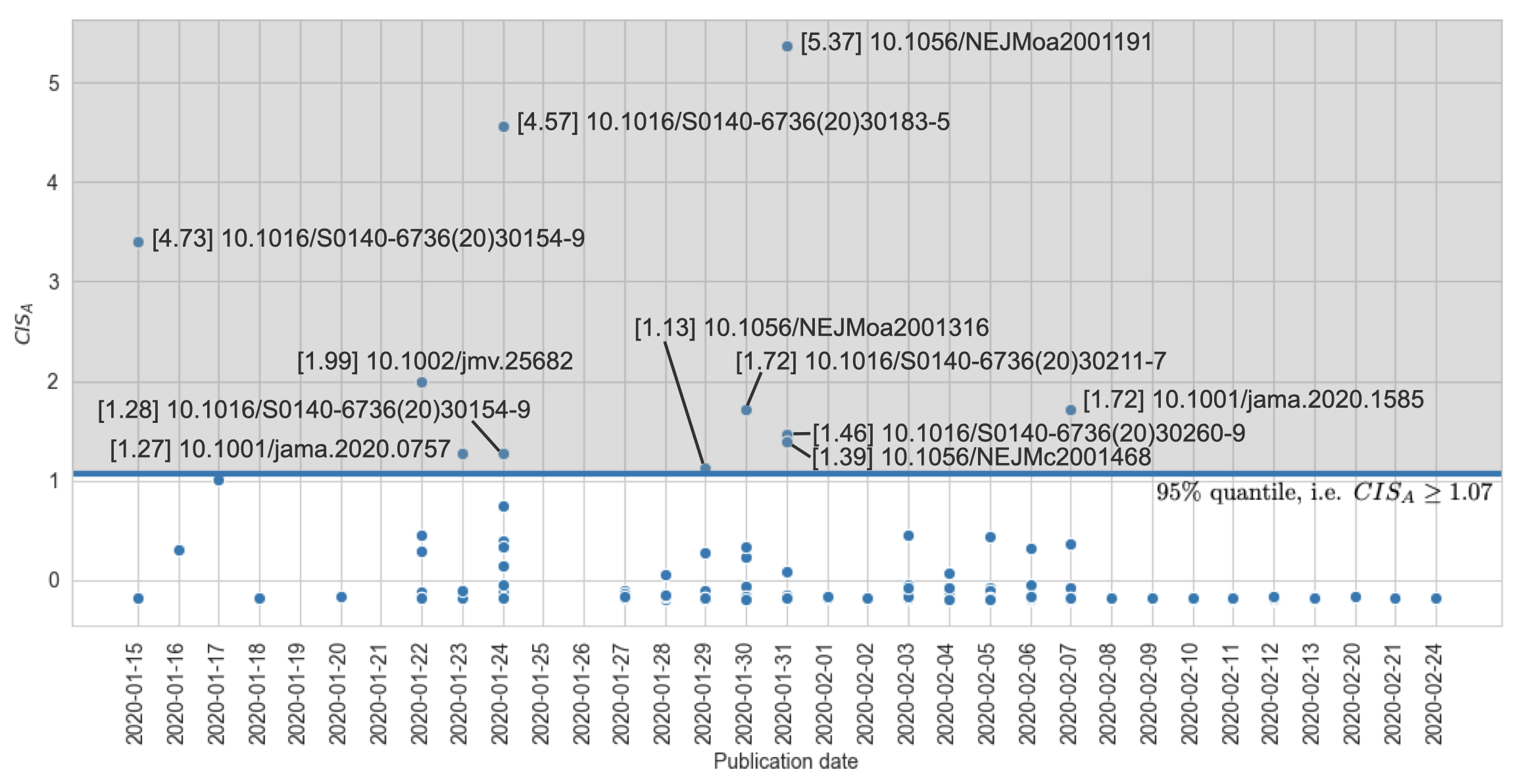}
    \caption{CIS computed on the set indicators $A$ limited to all altmetrics, i.e. captures, mentions, social media, and usage.}
    \label{fig:cis_altmetrics}
\end{subfigure}
\caption{Selection of papers based on the Comprehensive Impact Score (CIS) computed on citations and altmetrics separately.} 
\label{fig:cis_c_a}
\end{figure}

{\bf Comprehensive Impact Score.} 
On top of the different indicators we compute a {\em Comprehensive Impact Score} (CIS) for each paper in COVID-19-KG. CIS aims at providing a multi-dimensional and homogeneous view over indicators which are different in quantities and semantics, i.e. CIS represents a unifying score over heterogeneous bibliometric indicators. A paper CIS is computed by first standardising the values associated with each indicator category (e.g. number of social media mentions, number traditional citations, etc.) and then averaging the resulting values. We use $z$-scores (cf. Equation~\ref{eq:zeta}) for obtaining standard values and the arithmetic mean for the average. 
\begin{equation}
\label{eq:cis}
CIS(p) = \frac{\sum_{i \in I}z(p,i)}{|I|}
\end{equation}

\begin{figure}[!ht]
\begin{subfigure}[t]{\textwidth}
    \centering
    \includegraphics[scale=0.3]{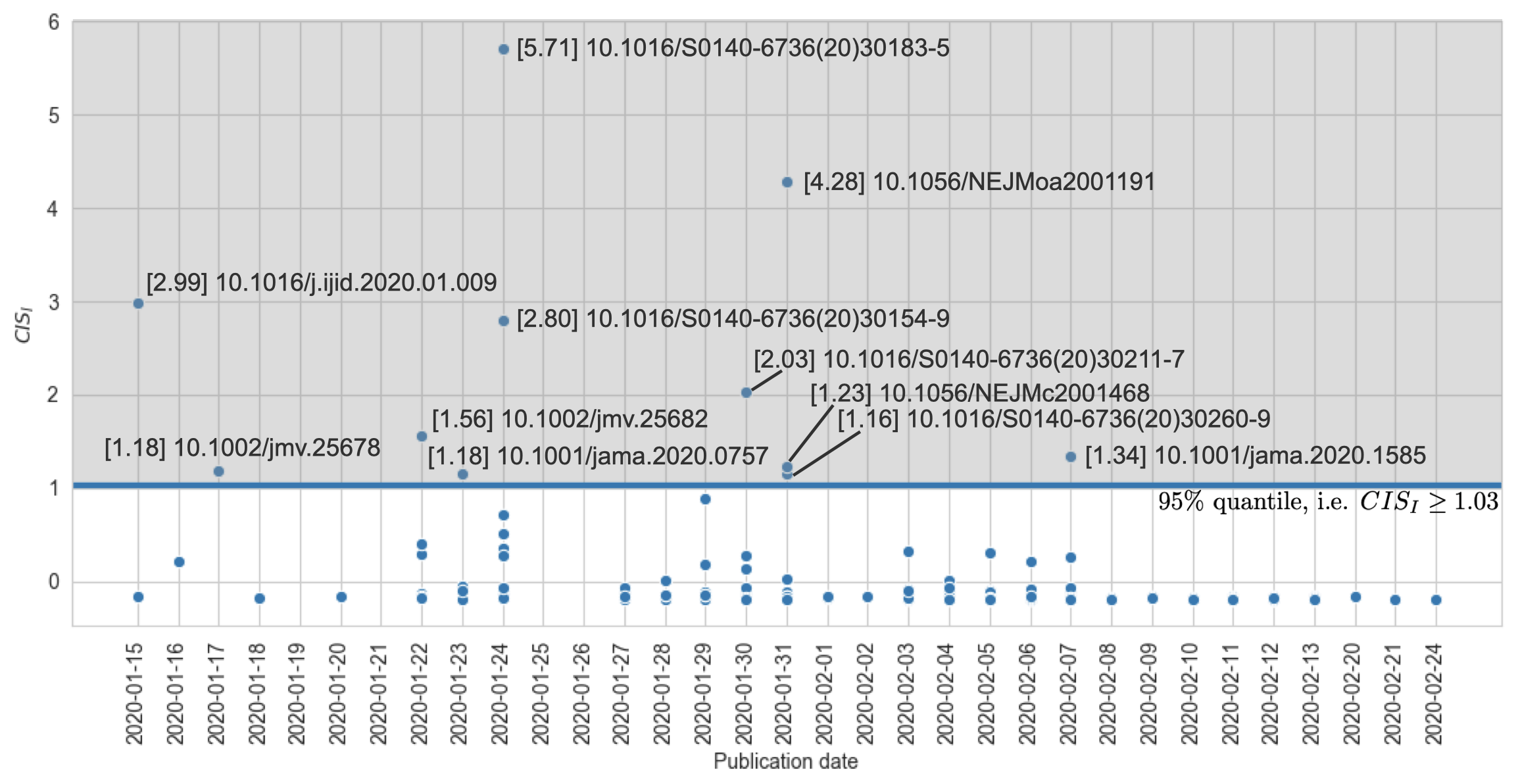}
    \caption{CIS computed on the all set of available indicators $I$ .}
    \label{fig:cis_full}
\end{subfigure}
\begin{subfigure}[t]{\textwidth}
    \centering
    \includegraphics[scale=0.3]{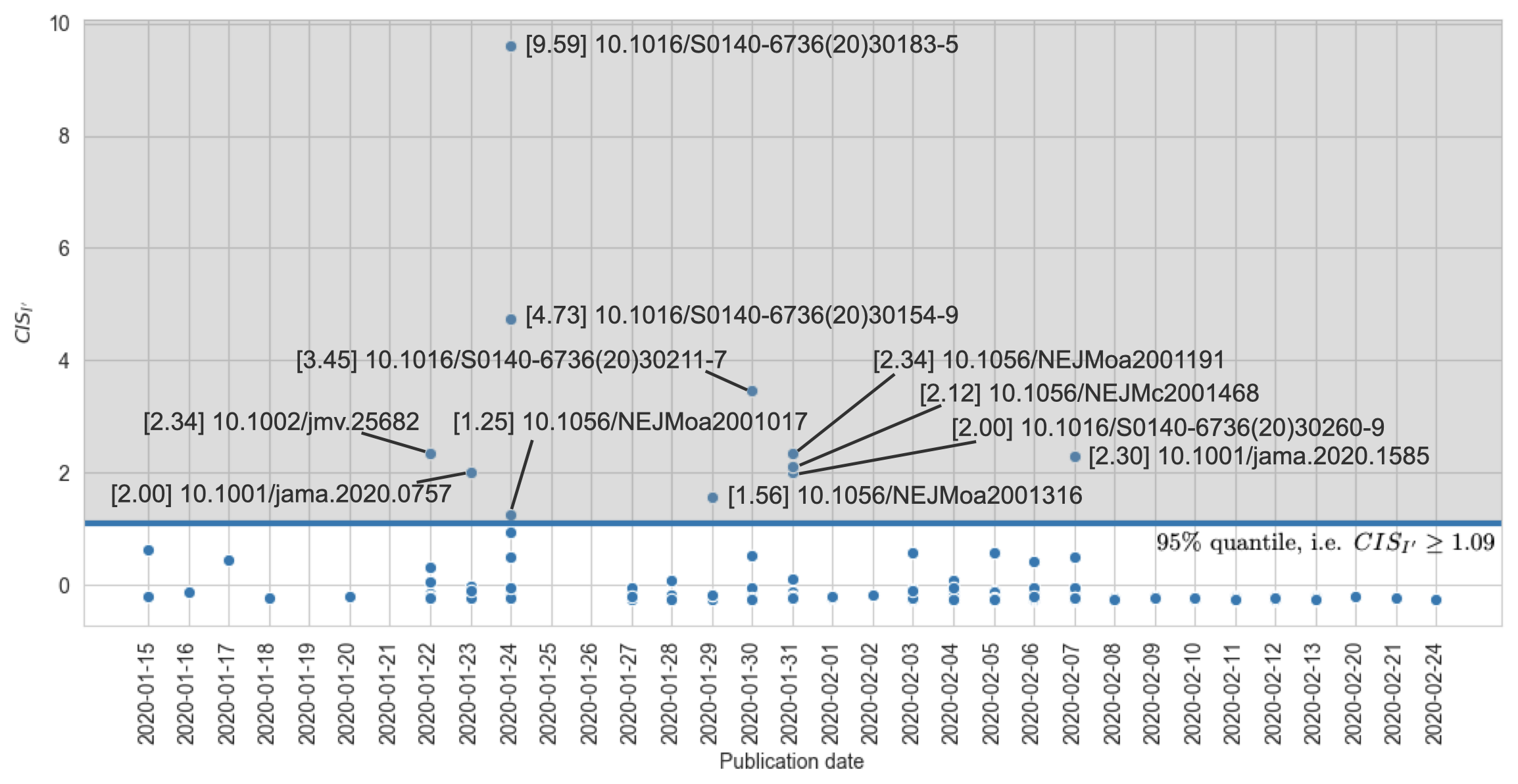}
    \caption{CIS computed on the set indicators $I'$ limited to citations, social media, and mentions.}
    \label{fig:cis_reduced}
\end{subfigure}
\begin{subfigure}[t]{\textwidth}
    \centering
    \includegraphics[scale=0.3]{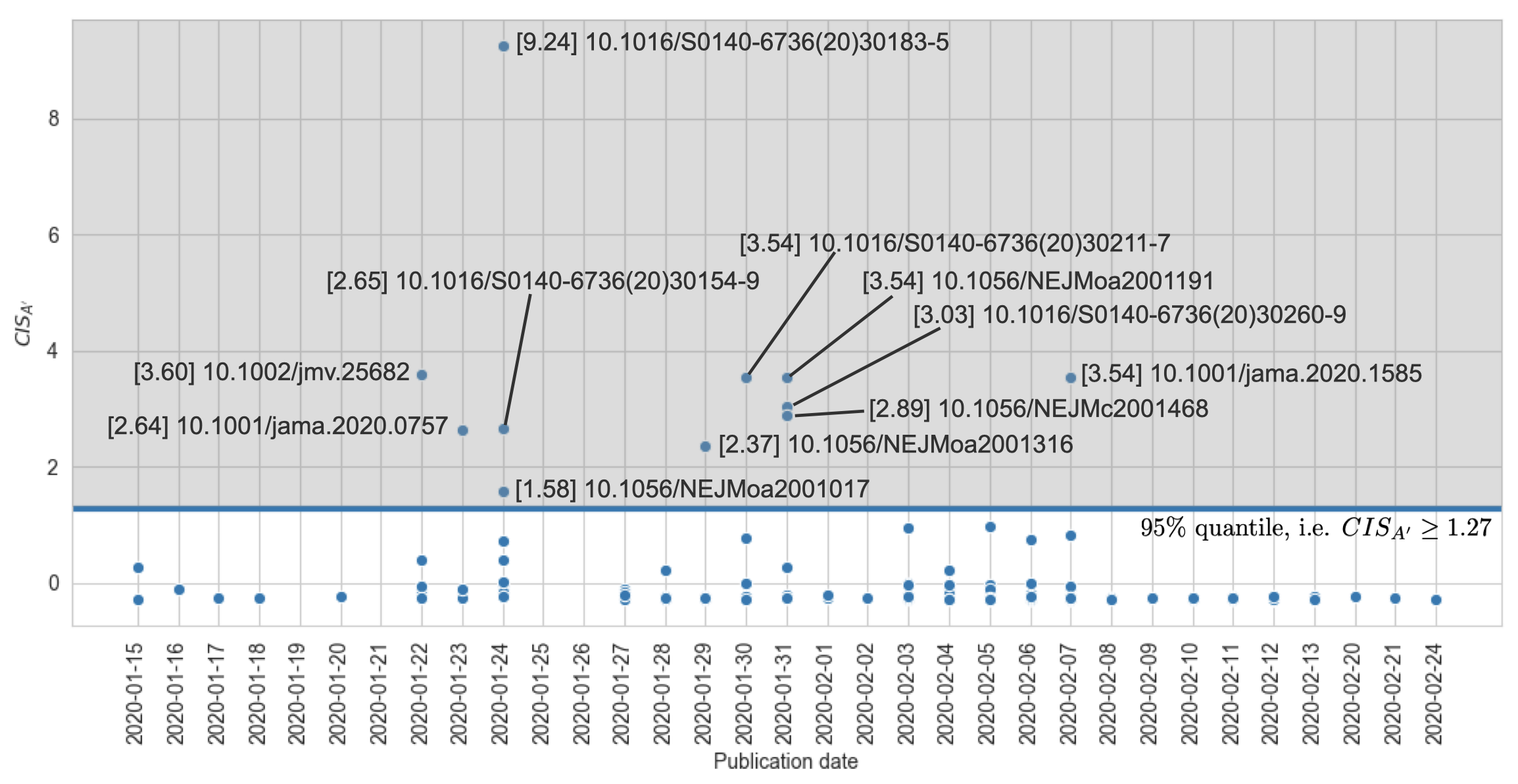}
    \caption{CIS computed on the set indicators $A'$ limited to social media and mentions.}
    \label{fig:cis_mentions_social_media}
\end{subfigure}
\caption{Selection of papers based on the Comprehensive Impact Score (CIS) computed on $I$, $I'$, and $A'$.} 
\label{fig:cis_alt}
\end{figure}

In Equation~\ref{eq:cis}: (i) $p$ is a paper that belongs to the set of available papers in COVID-19-KG; (ii) $i$ is an indicator that belongs to $I$, which, in turn, is the set of available indicators (e.g. citations, social media, etc.); and (iii) $z$ is the function for computing $z$-scores as defined in Equation~\ref{eq:zeta}. 
Finally, we compute the 95\% quantile on resulting CIS values. Again, the lower bound of the 95\% quantile is used as threshold value (i.e. $t$) for identifying candidate impactuful papers.


We use five distinct sets of indicators for investigating the role of altmetrics in the identification of impactful works (cf. {\em RQ2}) by means of CIS. Namely, those sets are: (i) citations only (i.e. $C$); (ii) the whole set of altmetrics available (i.e. $A$); (iii) citations and the whole set of altmetrics (i.e. $I$); (iv) citations, mentions, and social media (i.e. $I'$), which is the set comprising the categories with the highest correlation coefficients (cf. Section~\ref{sec:exp_behaviour}); and (v) mentions and social media (i.e. $A'$), which are the two categories of altmetrics with the highest correlation coefficients.
The lower bounds of the 95\% quantiles for CIS values computed over the five sets are: (i) $t_{CIS_{C}} \geq 0.27$; (ii) $t_{CIS_{A}} \geq 1.07$; (iii) $t_{CIS_{I}} \geq 1.03$, (iv) $t_{CIS_{I'}} \geq 1.09$; and (v) $t_{CIS_{A'}} \geq 1.27$. 
Figure~\ref{fig:cis_citations} shows the CIS values computed on the set of papers by taking into account citations only, i.e. $CIS_{C}$. Instead, Figure~\ref{fig:cis_altmetrics} shows the CIS values computed on the papers by taking into account the whole set of altmetrics, i.e. $CIS_{A}$. Both figures present papers distributed according to their publication date. Furthermore, in those figures the threshold $t$ is represented by the horizontal blue line the cut off discarded papers (i.e. those points below the threshold line) from selected ones (i.e. those points above the threshold line).  Similarly, Figure~\ref{fig:cis_full} shows the CIS values computed on the set of papers by taking into account citations and the whole set of altmetrics, i.e. $CIS_{I}$. Figure~\ref{fig:cis_reduced} shows the CIS values computed on the set of papers by taking into account citations, mentions, and social media, i.e. $CIS_{I'}$. Finally, Figure~\ref{fig:cis_mentions_social_media} shows the CIS values computed on the set of papers by taking into account mentions and social media, i.e. $CIS_{A'}$. 

\begin{figure}[t]
    \centering
    \includegraphics[scale=0.08]{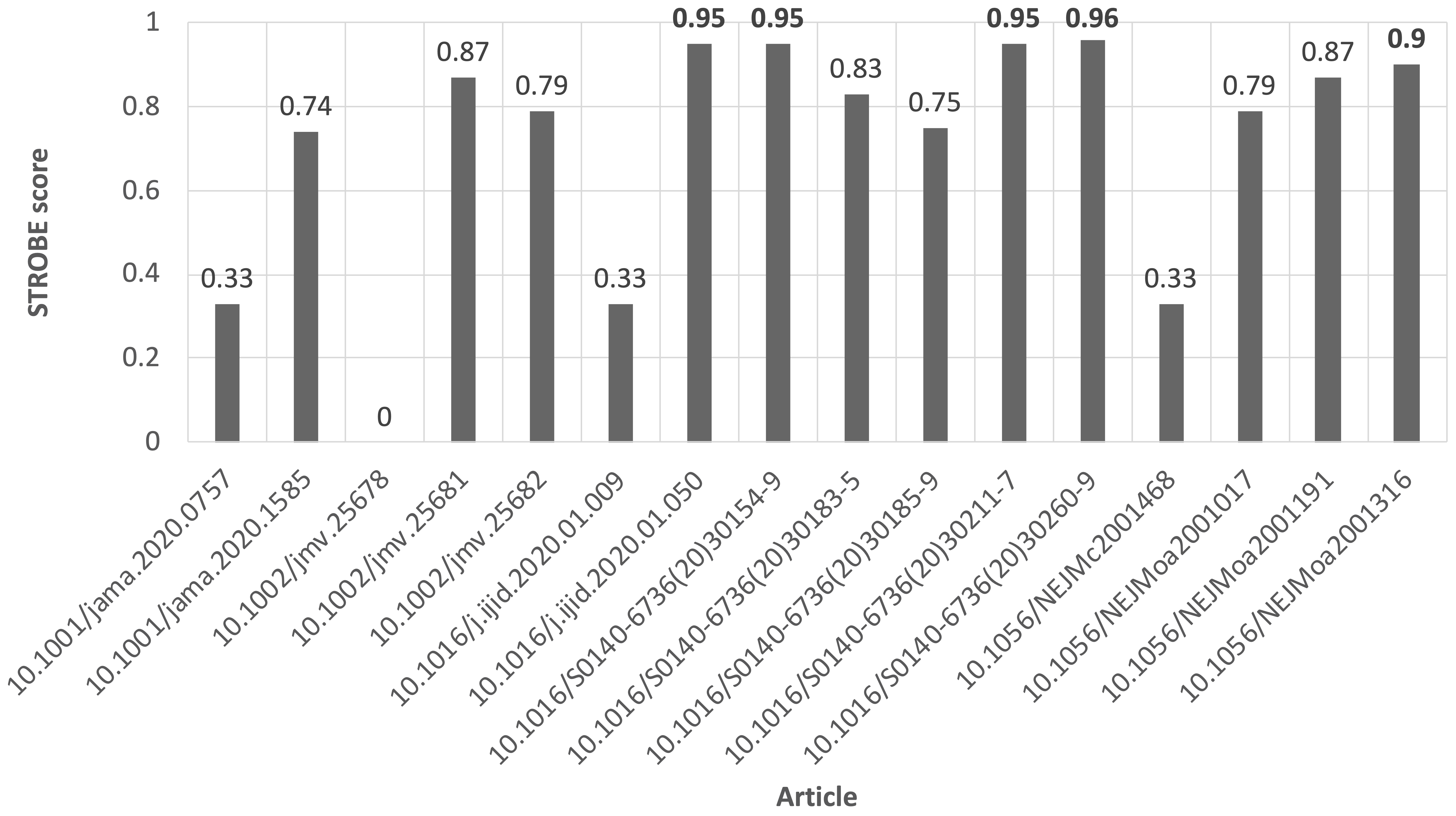}
    \caption{STROBE scores.}
    \label{fig:strobe}
\end{figure}

{\bf Expert-based assessment.} 
The articles automatically selected either by the geometric approach or the CIS one (cf. Table~\ref{tab:selected_cis}) are then assessed by means of the STrengthening the Reporting of OBservational studies in Epidemiology (STROBE) checklist\footnote{\url{https://www.strobe-statement.org/?id=available-checklists}}~\cite{Vandenbroucke2007}. STROBE is used to judge the quality of observational studies in meta-analyses. It consists of a checklist of 22 items, which relate to the title, abstract, introduction, methods, results and discussion sections of articles. Each item in the checklist is marked as checked only if the article addresses the assessment expressed by the item positively. Then, the quality of the paper is expressed as the ratio between the number of checked items and the total number of items in the checklist. We asked two human experts in epidemiology to carry out the STROBE assessment. They were unaware about the results obtained by either approaches, i.e. geometric selection or CIS-based. Both experts evaluated the articles independently in a first assessment step. Then, disagreements were resolved by discussion in a second step. Figure~\ref{fig:strobe} shows the STROBE scores assigned by the experts to the articles they were provided with. The full checklist filled by the experts is available on Zenodo\footnote{\url{https://doi.org/10.5281/zenodo.4026121}} as a CSV.
Instead, Table~\ref{tab:selected_cis} summarises the results by showing the articles selected by the different automatic approaches. The articles marked with the $+$ symbol are those who are selected by the automatic approaches and record strong STROBE scores, i.e. $strobe \geq 0.9$. We use 0.9 as threshold for this selection as it is the lower bound of the 4th quartile of the sample of the articles assessed by the humans. Hence, such a threshold is extremely selective in terms of quality. The articles marked with the $\bullet$ symbol are selected by the automatic approaches, but their associated STROBE score is lower than 0.9.

\begin{center}
\begin{table}[!ht]
\centering
\caption{Papers with their corresponding journal selected by using $CIS_{I}$ $CIS_{C}$, $CIS_{A}$, $CIS_{I}$, $CIS_{I'}$, $G_{c,s}$, $G_{c,m}$, and $G_{s,m}$.}
\label{tab:selected_cis}
\resizebox{\textwidth}{!}{ 
\begin{tabular}{l|r|c|c|c|c|c|c|c|c|c}
\centering
 {\bf Paper } & {\bf Journal } & $\mathbf{CIS_{C}}$ & $\mathbf{CIS_{A}}$ & $\mathbf{CIS_{I}}$ & $\mathbf{CIS_{I'}}$ & $\mathbf{CIS_{A'}}$ & $\mathbf{G_{c,s}}$ & $\mathbf{G_{c,m}}$ & $\mathbf{G_{s,m}}$\\ \hline \hline
 10.1001/jama.2020.0757 & JAMA & $\bullet$ & $\bullet$ & $\bullet$ & $\bullet$ & $\bullet$ & $\bullet$ &  & \\\hline
10.1001/jama.2020.1585 & JAMA &  & $\bullet$ & $\bullet$ & $\bullet$ & $\bullet$ &  &  & $\bullet$ \\\hline
10.1002/jmv.25678 & Med Virology & $\bullet$ &  & $\bullet$ &  &  &  &  & \\\hline
10.1002/jmv.25681 & Med Virology & $\bullet$ &  &  &  & & &  & \\\hline
10.1002/jmv.25682 & Med Virology &  & $\bullet$ & $\bullet$ & $\bullet$ & $\bullet$ &  &  & $\bullet$ \\\hline
10.1016/j.ijid.2020.01.009 & IJID & $\bullet$ & $\bullet$ & $\bullet$ &  & &  &  & \\\hline
{\bf 10.1016/j.ijid.2020.01.050} & IJID & {\bf + } &  &  &  &  &  & & \\\hline
{\bf 10.1016/S0140-6736(20)30154-9} & Lancet & {\bf + } & {\bf + } & {\bf + } & {\bf + } &  {\bf + } & {\bf + } & {\bf + } & {\bf + }\\\hline
10.1016/S0140-6736(20)30183-5 & Lancet & $\bullet$ & $\bullet$ & $\bullet$ & $\bullet$ & $\bullet$ & $\bullet$ & $\bullet$ & $\bullet$ \\\hline
10.1016/S0140-6736(20)30185-9 & Lancet & $\bullet$ &  &  &  & &  &  & \\\hline
{\bf 10.1016/S0140-6736(20)30211-7} & Lancet & {\bf + } & {\bf + } & {\bf + } & {\bf + } & {\bf + }& {\bf + } & {\bf + } & {\bf + }\\\hline
{\bf 10.1016/S0140-6736(20)30260-9} & Lancet &  & {\bf + } & {\bf + } & {\bf + } & {\bf + } &  &  & {\bf + }\\\hline
10.1056/NEJMc2001468 & NEJM & $\bullet$ & $\bullet$ & $\bullet$ & $\bullet$ & $\bullet$ & $\bullet$ & $\bullet$ & $\bullet$ \\\hline
10.1056/NEJMoa2001017 & NEJM & $\bullet$ &  &  & $\bullet$ & $\bullet$ & $\bullet$ & $\bullet$ & $\bullet$ \\\hline
10.1056/NEJMoa2001191 & NEJM &  & $\bullet$ & $\bullet$ & $\bullet$ & $\bullet$ & &  & $\bullet$ \\\hline
{\bf 10.1056/NEJMoa2001316} & NEJM &  & {\bf + } &  & {\bf + } & {\bf + } & &  & \\\hline
\end{tabular}
}
\end{table}
\end{center}

\section{Discussion}
\label{sec:discussion}
{\bf Behavioural perspective.} The density estimation based on Gaussian kernels, i.e. KDE, shows that for COVID-19-KG all categories provide sparse indicators. However, if we analyse the density curves for individual metrics we observe clear patterns that characterise each indicator category uniquely. On one hand, for the scope of this work, we investigate those patterns in order to understand how indicator categories behave by using the data coming from COVID-19-KG. On the other hand, it is worth saying that patterns from KDE are specifically suitable for working in scenarios in which inference is required. For example, an algorithm might leverage learned patterns for implementing a classic binary classification task. This task might require to identify relevant papers from other samples with similar characteristics (e.g. a similar time-window). A typical classification task might distinguish papers according to the relevant/not-relevant dichotomy.
Accordingly, as future work, it would be interesting to associate KDE probability with impact categories, e.g. those emerging from the geometric space analysis or the CIS one. The KDE based on $z$-scores shows that the density curves are mostly overlapped with each other, both at category and source level. Thus, we observe a similar citational behaviour once the indicator values are standardised, as a shared pattern clearly emerges from their density curves. However, it is fairly evident that the KDE based on $z$-scores flattens differences with regards to both the indicator values and indicator semantics. For example, social media counts range from 0 to 45,197 with 1,250.34 and 36.5 as mean and median, respectively, while citations ranges from 0 to 82 with 1.63 and 0 as mean and median, respectively (cf. Table~\ref{tab:altmetrics_stats}). Numerical differences are captured by KDE if computed on individual categories (cf. Figure~\ref{fig:density}), but different semantic flavours among indicators are not. 

We do not design any schema that formally captures the meaning of each indicator (e.g. an ontology like those proposed by~\cite{Darcus2008} or~\cite{Shotton2010}). Nevertheless, we investigate if any correlation among any pair of indicators within the category-metric-source hierarchy can be interpreted as similarity in usage. The correlation analysis suggests that citations, social media, and mentions identify a cluster of indicators that is used with a certain degree of consistency by citing entities. This finding confirms the outcomes of~\cite{Kousha2020}. Furthermore, the correlation analysis performed by properly selecting samples of articles by removing zero values as indicators suggests that the correlation coefficients are not inflated by zero values. Hence, the correlation among citations, social media, and mentions is not obtained by chance. This suggests that citations, social media, and mentions exhibit a similar behaviour in our sample about COVID-19 litereture, despite being heterogeneous and identifying different categories of indicators~\cite{Haustein2016}. We leverage this similarity in order to investigate {\em RQ2}, that is, understanding how to use altmetrics for detecting impact. In~\cite{Nuzzolese2019} we performed a similar analysis, and we recorded good correlation between citations and usage (i.e. the Mendeley readers). This different result should not be misinterpreted. We are persuaded that the outcomes presented in~\cite{Nuzzolese2019} are valid. Simply, they are obtained on a data sample which is inherently different from COVID-19-KG. Indeed, the data sample used in~\cite{Nuzzolese2019} is much larger and wider in terms of papers contained (i.e. 833,116 against 212 in COVID-19-KG) and the time-window encompassing publication dates (i.e. many years against a month in COVID-19-KG), respectively. Accordingly, the research questions leading this work are different (cf. {\em RQ1} and {\em RQ2}). As a matter of fact, the limited time-window is a mandatory requirement inherently related to the limited scientific literature we have since COVID-19 first appeared in human history (approximately December 2019). Based on our results we are happy to record social media (i.e. Twitter and Facebook) and mentions (limited to News and Blogs only) as valid tools for fast and early scholarly communication. This follows the visionary intuition of the altmetrics manifesto\footnote{\url{http://altmetrics.org/manifesto/}}~by \cite{Priem2012}. The moderate occurrence of traditional citation counts with a certain positive (i.e. $r>0.6$) and statistically significant (i.e. $p<0.01$) correlation with altmetrics (i.e. social media and mentions) is, instead, a bit surprising. The statistics about usage (cf. Table~\ref{tab:altmetrics_stats}), along with the results recorded for the KDE and the correlation analysis entitle us to claim that tweets on Twitter, shares and likes on Facebook, mentions on news and blogs, and citations in academic literature tracked by Scopus, are the channels that have been mostly used for early scholarly communication about COVID-19. This addresses {\em RQ1}. 

{\bf Impact analysis.} The selection method based on the geometric space shows that mentions and social media (i.e. $G_{s,m}$) are more reliable than other pairs, when they are used together as axes of the Cartesian plane meant for positioning papers geometrically (cf. Table~\ref{tab:selected_cis}). In fact, by using them, we record a set of 9 candidate papers, among which 3 are validated as extremely qualitative by human experts. On the contrary, $G_{c,s}$ and $G_{c,m}$ record 6 and 5 candidate papers respectively. Additionally, in the set of candidate papers identified by $G_{c,s}$ and $G_{c,m}$ only 2 are validated by human experts. If we assess the selection based on the impact of the journals the papers have been published in, then we record good evidence about quality. In fact, most of the candidate papers appeared in the New England Journal of Medicine and The Lancet, which are in the top-5 journal ranking on medicine according to SCImago\footnote{\url{https://www.scimagojr.com/journalrank.php?category=2701}}, with an SJR of 19.524 and 15.871, respectively.
With regards to the exhaustiveness, the selection of candidate impactful papers is, in our opinion, an exploratory search task. According to~\cite{White05}, exploratory search tasks are typically associated with undefined and uncertain goals. This means that identifying all possible impactful papers is nearly impossible. Hence, dealing with sub-optimal exhaustiveness is the practice in scenarios like these due to the inherent nature of the search problem.  

The selection based on the Comprehensive Impact Score (CIS) overcomes the limitation of a two-dimensional space introduced when defining a selection method based on a Cartesian plane. Indeed, CIS is a multi-dimensional selection tool which is customisable in terms of the indicators used for performing the analysis. It is fairly evident (cf. Table~\ref{tab:selected_cis}) that $CIS_{C}$, $CIS_{A}$, $CIS_{I}$, $CIS_{I'}$, and $CIS_{A'}$ share most of the papers identified by $G_{c,s}$, $G_{c,m}$, and $G_{s,m}$. All five CIS-based selections extend the set of selected articles returned by $G$ with 6 additional works (cf. Table~\ref{tab:selected_cis}) published in The Lancet, the International Journal of Infectious Diseases ($SJR=1.456$), JAMA ($SJR=7.477$), the Journal of Medical Virology ($SJR=0.966$), and the New England Journal of Medicine. Among those additional papers only the one appeared in the Journal of Medical Virology is debatable both for the journal impact (i.e. $SJR=0.966$) and its scientific relevance. In fact, in this paper the authors claim that two type of snakes, which are common in Southeastern China, are the intermediate hosts responsible for the ``cross‐species'' transmission of SARS-CoV-2 to humans. Subsequent genomic studies\footnote{\cite{Andersen2020} is not part of the COVID-19 sample as it has been published on March 17th 2020, while the upper bound of time-window of the COVID-19 sample is February 24th 2020.}~\cite{Andersen2020} confirm cross‐species transmission, but they refute the theory of snakes being the intermediate hosts. However, the paper reporting the theory of snakes being the intermediate hosts has been largely (i) retweeted, shared, and liked on different social networks, and (ii) discussed and reported by many international newspapers worldwide. Thus it contributed to the massive ``infodemic''\footnote{\url{https://www.who.int/docs/default-source/coronaviruse/situation-reports/20200202-sitrep-13-ncov-v3.pdf}} about COVID-19, i.e. an over-abundance of information, either accurate or not, which makes it hard for people to find trustworthy sources and reliable guidance when they need it~\cite{Zarocostas2020}. This infodemic is captured by altmetrics that, by design, are fed by on-line tools and platforms. As a matter of fact, the paper about the theory of snakes being the intermediate hosts is selected among the candidates only by $G_{s,m}$, where both axes are altmetrics, and not by $G_{c,s}$ and $G_{c,m}$, where traditional citations are taken into account. This suggests a twofold speculation: (i) the scientometric community should handle altmetrics very carefully as they may lead to unreliable and debatable results; (ii) altmetrics are promising tools not only for measuring impact, but also to make unwanted scanarios (e.g. infodemic) emerge from the knowledge soup of scientific literature. We opt for the second. Nevertheless, further and more focused research is needed.

Finally, the expert-based quality assessment records that the CIS based on the whole set of altmetrics (i.e. $CIS_{A}$), the one based on mentions and social media (i.e. $CIS_{A'}$), and the one based on mentions and social media combined with traditional citations (i.e. $CIS_{I'}$) provide the most articles (i.e. 5 works as reported in Table~\ref{tab:selected_cis}) that are marked as extremely qualitative by human experts. Additionally the sets of articles with $STROBE \geq 0.9$ identified by $CIS_{A'}$ and $CIS_{I'}$ are equivalent. This suggests that, at very early stages of the scientific production, altmetrics might be used effectively as tools for measuring impact regardless of traditional citations. Furthermore, in our experiments (i) tweets on Twitter, (ii) shares, likes and comments on Facebook, and (iii) mentions on blogs and news are, besides the others, the sources of altmetrics that convey the most impact, i.e. $CIS_{A'}$. This addresses {\em RQ2}. It is worth clarifying that altmetrics are heterogeneous in semantics as they are computed over a broad set of sources which are different in terms of users and purposes~\cite{Haustein2016}. We do not claim that CIS is effective for compounding citations and a range of altmetrics, hence flattening a multi-dimensional scenario which is associated with the citational context of scientific works. On the contrary, we claim that 
CIS provides a fairly simple unifying view based on those indicators that share similarities. This is the reason why we perform the behavioural analysis based on density and correlation among indicators.

\section{Conclusions and future work}
\label{sec:conclusion}
In this work we investigate altmetrics and citation count as tools for detecting impactful research works in quasi-zero-day time-windows, as it is for the case of COVID-19. COVID-19 offers an extraordinary real-world case-study for understanding inherent correlation among impact and altmetrics. As mentioned in Section~\ref{sec:intro}, for the first time in history, humankind is facing a pandemic, which is described, debated, and investigated in real time by the scientific community via conventional research venues (i.e. journal papers), and social and on-line media. The latters are the natural playground of altmetrics. Our case-study relies on a sample of 212 scientific papers on COVID-19 collected by means of a literature review. Such a literature review is based on a two-stage screening process used to assess the relevance of studies on COVID-19 appeared in literature from January 15th 2020 to February 24th 2020. This sample is used for constructing a knowledge graph, i.e. COVID-19-KG, modelled in compliance with the Indicators Ontology (i.e. I-Ont). COVID-19-KG is the input of our analysis aimed at investigating (i) behavioural characteristics of altmetrics and citation count and (ii) possibible approaches for using altmetrics along with citation count for automatically identifying candidate impactful research works in COVID-19-KG. We find moderate correlation among traditional citation count, citations on social media, and mentions on news and blogs. This suggests there is a common intended meaning of the citational acts associated with these indicators. Additionally, we define the {\em Comprehensive Impact Score} (CIS) that harmonises different indicators for providing a multi-dimensional impact indicator. CIS shows to be a promising tool for selecting relevant papers even in a tight observation window. Possible future work include the use of CIS as a feature for predicting the results of evaluation procedures of academics as presented in works like~\cite{Poggi2019}. Similarly, further investigation is needed to mine the rhetorical nature of citational acts associated with altmetrics. The latter is a mandatory step for building tools such as~\cite{Ciancarini2014} and~\cite{Peroni2020}. More ambitiously, future research focused on altmetrics and citations should go in the direction envisioned by~\cite{Gil2014} and~\cite{Kitano2016}, thus
contributing to a new family of artificial intelligence aimed at achieving autonomous {\em discovery science}.

\bibliographystyle{aps-nameyear}      
\bibliography{references-new}                 
\nocite{*}

\end{document}